\documentclass{aa}

\usepackage{graphicx}
\usepackage{adjustbox}
\usepackage{txfonts}
\usepackage{url}
\usepackage{relsize}
\usepackage{color}
\usepackage{amstext}
\usepackage{amsmath}

\begin{document}

\newcommand \bprime {\backprime\hspace{-.11em}      }   %  backstroke
\newcommand \xprime {\prime\hspace{-.25em}\backprime}   %  cross
\def\dzero{{(0)}}
\def\dtwo{{\prime\prime}}
\def\dthree{{\prime\prime\prime}}
\def\dfour{{(4)}}
\def\dfive{{(5)}}
\def\dsix{{(6)}}
\def\dn{{(n)}}

\def\simlt{\mathrel{\rlap{\lower 3pt\hbox{$\sim$}}\raise 2.0pt\hbox{$<$}}}
\def\simgt{\mathrel{\rlap{\lower 3pt\hbox{$\sim$}} \raise 2.0pt\hbox{$>$}}}

\def\lsim{\,\lower2truept\hbox{${< \atop\hbox{\raise4truept\hbox{$\sim$}}}$}\,}
\def\gsim{\,\lower2truept\hbox{${> \atop\hbox{\raise4truept\hbox{$\sim$}}}$}\,}

\def\beq{\begin{equation}}
\def\eeq{\end{equation}}

         \title{On the fast computation of the observer motion effects induced on monopole frequency spectra for tabulated functions}   

    \author{T. Trombetti\inst{1}\fnmsep\thanks{e-mail:trombetti@ira.inaf.it}
                     \and
           C. Burigana\inst{1,2} \and M. Tucci\inst{3} \and L. Toffolatti\inst{4,5}
           }
    \institute{INAF, Istituto di Radioastronomia, Via Piero Gobetti 101, I-40129 Bologna, Italy
    \and
    INFN, Sezione di Bologna, Via Irnerio 46, I-40127 Bologna, Italy
    \and
   Department of Astronomy, University of Geneva, ch. d’\'{E}cogia 16, CH--1290 Versoix, CH
    \and
    Departamento de F\'{i}sica Universidad de Oviedo, C. Federico Garc\'{i}a Lorca 18, E--33007 Oviedo, Spain
    \and
    Instituto Universitario de Ciencias y Tecnolog\'{i}as Espaciales de Asturias (ICTEA), C. Independencia 13, E-33004 Oviedo, Spain 
    }

   \date{Received ...; accepted ...}

 \abstract
 {Various methods have been studied to compute the boosting effects produced by the observer peculiar motion that modifies and transfers to higher multipoles the isotropic monopole frequency spectrum of the cosmic background radiation. Explicit analytical solutions for the spherical harmonic expansion coefficients were presented and applied to different types of background spectrum, strongly alleviating the computational effort needed for accurate theoretical predictions. The frequency spectra at higher multipoles are inherently led by higher order derivatives of the monopole spectrum. Provided that it can be well described by analytic or semi-analytic functions, the computation of its transfer is not affected by numerical instabilities when evaluated at the needed level of numerical accuracy. Instead, monopole frequency spectra described by tabulated functions are computed with a relatively poor frequency resolution in comparison with the Doppler shift, thus requiring to interpolate the tabular representation. They are also affected by uncertainties related to intrinsic inaccuracies in the modelling or in the related observational data as well as to limited accuracy in their numerical computation. These uncertainties propagate and 
increase with the derivative order, possibly preventing a trustworthy computation of the transfer to higher multipoles and of the observed monopole.}
 {We study methods to filter the original function or its derivatives and the multipole spectra, to mitigate numerical instabilities and derive reliable predictions of the harmonic coefficients for different cosmic background models.}
{From the analytical solutions and assuming that the monopole spectrum can be expanded in Taylor's series, we derive explicit expressions for the harmonic coefficients up to the multipole $\ell_{\rm max} = 6$ in terms of monopole spectrum derivatives. We then consider different low-pass filters: pre-filtering in Fourier space of the tabular representation; filtering in both real and Fourier space of the numerical derivatives; interpolation approaches; a dedicated method based on the boosted signal amplification and deamplification. We study the quality of these methods on suitable analytical approximations of the tabulated functions, possibly polluted with simulated noise, for their further application to the tabulations.}
 {We consider two very different types of monopole spectra superimposed to the cosmic microwave background, the (smooth) extragalactic sources microwave background signal from radio loud active galactic nuclei and the (feature-rich) redshifted 21cm line, and present our results in terms of spherical harmonic coefficients. Their direct prediction can be noisy at $\ell > 1$ or, depending on the uncertainty level, even at $\ell \le 1$. Without assuming a functional form for the extragalactic background spectrum, the Gaussian pre-filtering coupled to the sequential real space filtering of derivatives allows us to derive accurate predictions up to $\ell \sim 6$, while a log-log polynomial representation, appropriate on several decades, gives accurate solutions at any $\ell$. Instead, it is difficult to characterise the 21 cm line model variety, so it is relevant to work without assumptions about the underlying function. Typically, the pre-filtering provides accurate predictions up to $\ell \simeq 3$ or $4$, while the further sequential filtering of the derivatives or the boosting amplification and deamplification method improves the results up to $\ell = 4$, also allowing reasonable estimations of the spectrum at higher $\ell$.}
 {The proposed methods can significantly extend the range of realistic cosmic background models manageable with a fast computation, beyond the cases characterised a priori by analytical or semi-analytical functions, requiring only an affordable increase in computational time compared to the direct calculation via simple interpolation.} 

   \keywords{cosmic background radiation -- diffuse radiation -- methods: analytical}
      
      \titlerunning{Computation of observer motion effect for tabulated background spectra}

   \maketitle

%-------------------------------------------------------------------

\section{Introduction}
\label{sec:intro}

The observer peculiar motion with respect to the cosmic background radiation produces boosting effects that modify and transfer to higher multipoles the isotropic monopole frequency spectrum of the background, which can therefore be investigated through the frequency dependence of the signal variation in the sky. Various methods have been studied to compute these effects for different kinds of background signals and to explore the perspectives of such differential approaches in various observational frameworks \citep[e.g.,][]{1981A&A....94L..33D,2015ApJ...810..131B,2016JCAP...03..047D,2017PhRvL.118o1301S,2018JCAP...04..021B,2018MNRAS.477.4473M,2018ApJ...866L...7D,2019A&A...631A..61T}.

The use of the dipole spectrum as an additional probe of the background monopole spectrum was proposed more than 40 years ago, but only recently this approach has been deeply investigated in the context of future experiments as a way to circumvent the difficulties represented by the absolute calibration and to deal with the foreground impact that could be less relevant in dipole than in monopole analyses.
The need for relative and inter-frequency calibration, common to both absolute and differential approaches, although in principle less critical than the absolute one, represents a key aspect. In the context of future cosmic microwave background (CMB) missions, including the detection and characterization of spectral distortions in their scientific aims, the impact in dipole analyses of potential residuals from imperfect calibration and the conceptual design of instrument and measurement techniques for the final output calibration have been already addressed in \citet{2018JCAP...04..021B} and in \citet{2018MNRAS.477.4473M}, respectively. 

While the transfer of the isotropic monopole spectrum to multipole spectra beyond the dipole or quadrupole could be the subject of future dedicated observational studies, we focus here on the fast computation of the coefficients of the spherical harmonic expansion for generic background functions.

Analytical solutions of a system of linear equations to explicitly compute the terms of spherical harmonic expansion have been recently presented and applied to different types of background spectrum and their combinations, significantly alleviating the computational effort needed for accurate theoretical predictions, also including high order contributions \citep{2021A&A...646A..75T}.
The frequency spectra at higher multipoles are inherently led by higher order derivatives of the monopole frequency spectrum. Provided it can be exhaustively described by analytic or semi-analytic functions and the calculation is performed at the needed level of numerical accuracy, the computation of the frequency spectra at higher multipoles is not affected by numerical instabilities.
The situation is very different for monopole frequency spectra described by tabulated functions. They are typically computed with a relatively poor resolution in frequency, $\nu$, in comparison with the order of the Doppler shift in frequency, $\delta \nu / \nu \sim \beta = {\rm v} / c$, ${\rm v}$ and $c$ being respectively the speed of observer and of light, thus requiring to interpolate the tabular representation. Furthermore, they are affected by uncertainties related to intrinsic inaccuracies in astrophysical modelling or in related ingested observations or to limited accuracy in their numerical computation. These inaccuracies propagate and relatively increase,
possibly causing significant instabilities, with the derivative order. According to the quality of the tabulated monopole frequency spectrum,
these inaccuracies and instabilities may in principle prevent a robust and accurate computation of its transfer to higher multipoles or even at $\ell \le 1$.
The scope of this paper is to identify approaches that can be applied to derive reliable predictions of the harmonic coefficients for a wide range of background models.
 
In Sect. \ref{sec:theoframe} we introduce the adopted formalism. Starting from the analytical solutions for the spherical harmonic coefficients and assuming that the monopole frequency spectrum can be expanded in Taylor's series,
we derive explicit expressions for the spherical harmonic coefficients in terms of monopole frequency spectrum derivatives and discuss their scaling relations relevant for the following sections.
Among the plethora of possible background models, we focus on two classes of monopole spectra (Sect. \ref{sec:monmod}) usually represented by tabulated functions, namely the (smooth) extragalactic sources microwave background (ESMB) signal from radio loud active galactic nuclei (AGN) and the (feature-rich) redshifted 21cm line that are superimposed to the CMB spectrum. These two classes of models are very different in the size of the frequency range in which they are relevant and in the spectral shape, thus offering the possibility to test the developed methods under very different conditions.
In Sect. \ref{sec:instability} we show the type of inaccuracy and instability appearing in the computation of the transfer of the monopole frequency spectrum to the higher multipoles applying directly the analytical solutions.
We study the effects of these approaches to frequency spectra at different multipoles considering first suitable analytical approximations of the tabular representations, 
possibly polluted with simulated noise in the range of the quoted uncertainties of the models, and then tabulated functions. The results are presented in terms of spherical harmonic coefficients.
In Sect. \ref{sec:filt_der} we discuss methods to filter the original tabulated function or its derivatives and, ultimately, the higher multipole spectra to mitigate numerical instability. 
We consider different filters, from pre-filtering in Fourier space of the original tabulated background function to filtering in both real and Fourier space of the numerical derivatives, from local and, where feasible, global interpolation approaches to a dedicated method of amplification and deamplification of the boosting.
The above methods are applied to the two considered classes of monopole spectra in Sect. \ref{sec:results}, where the main results are shown.
The comparison among the results found for the different filtering methods applied to these classes of monopole spectra, the guidelines to help the identification of the best types of treatments, the discussion of the required computational time and the main conclusions are given in Sect. \ref{sec:conclu}.

\section{Theoretical framework}
\label{sec:theoframe}

The Compton-Getting effect \citep{1970PSS1825F}, based on the Lorentz invariance of the photon distribution function, $\eta(\nu)$, allows to describe how the peculiar velocity of an observer impacts the frequency spectrum on the whole sky. At the frequency $\nu$, the observed signal in equivalent thermodynamic temperature, $T_{\rm th} (\nu) ={(h\nu/k)} / {\ln(1+1/\eta(\nu))}$, is
\begin{align}\label{eq:eta_boost}
T_{\rm th}^{\rm BB/dist} (\nu, {\hat{n}}, \vec{\beta}) & =
\frac{xT_{0}} {{\rm{ln}}(1 + 1 / (\eta(\nu, {\hat{n}}, \vec{\beta}))^{\rm BB/dist}) } \nonumber
\\ & = \frac{xT_{0}} {{\rm{ln}}(1 + 1 / \eta(\nu')) }\, ,
\end{align}
\noindent
where
\begin{equation}\label{eq:etaboost}
\eta(\nu, {\hat{n}}, \vec{\beta}) = \eta(\nu') \, ,
\end{equation}
\noindent
with
\begin{equation}\label{eq:nuboost}
\nu' = \nu (1 - {\hat{n}} \cdot \vec{\beta})/(1 - \beta^2)^{1/2} \, .
\end{equation}
\noindent
In the above formulas, ${\hat{n}}$ is the sky direction unit vector, $\vec{\beta} = \vec{{\rm v}} / c$ is the observer velocity, $x=h\nu/(kT_r)$ is the CMB redshift invariant dimensionless frequency, $T_r=T_0(1+z)$ is the CMB redshift dependent effective temperature, $k$ and $h$ are the Boltzmann and Planck constants, and the notation `BB/dist' indicates a blackbody spectrum or any type of spectral distortion \citep{2018JCAP...04..021B}. 
Here $T_0$ is the current CMB effective temperature in the blackbody spectrum approximation such that
$aT_0^4$ gives the current CMB energy density with $a = 8\pi I_3 k^4 / (hc)^3$, $I_3 = \pi^4/15$. From the joint analysis of the data from the Far Infrared Absolute Spectrophotometer (FIRAS)
on board the Cosmic Background Explorer (COBE) and from the Wilkinson Microwave Anisotropy Probe (WMAP),
\cite{2009ApJ...707..916F} derived $T_0 = (2.72548 \pm 0.00057)$\,K.

Expanding in spherical harmonics coefficients Eq. \eqref{eq:eta_boost} gives:
\begin{equation}\label{eq:harm}
T_{\rm th}^{\rm BB/dist} (\nu, \theta, \phi, \beta) =
\sum_{\ell=0}^{\ell_{\rm max}} \sum_{m=-\ell}^{\ell} a_{\ell,m} (\nu, \beta) Y_{\ell,m}(\theta, \phi) \, .
\end{equation}
\noindent
To simplify the problem, we choose a reference system such as to maintain the dependence on the colatitude $\theta$ and make that on the longitude $\phi$ vanish, i.e. with the $z$ axis parallel to the observer velocity. This means that only the coefficients $a_{\ell,m} (\nu, \beta)$ with $m=0$ do not vanish. The $a_{\ell,0} (\nu, \beta)$ amplitude decreases as $\beta^{\ell \cdot p}$ at increasing multipole, $\ell$, with $p \approx 1$ (for a BB, $p=1$ and $a_{\ell,0} (\nu, \beta) = a_{\ell,0} (\beta)$).

\subsection{Structure of analytical solutions}
\label{sec:sol_struct}

Fixing $\ell_{\rm max}$ to a specific value in Eq. \eqref{eq:harm}, it is possible to identify a system of linear equations in the $N$ unknowns $a_{\ell,0} (\nu, \beta)$, being $N = \ell_{\rm max}+1$ the number of sky directions, evaluating the signal $T_{\rm th}^{\rm BB/dist}$ through Eq. \eqref{eq:eta_boost}. From \cite{2020A&A...641A...1P} we know that $\beta = (1.2336 \pm 0.0004) \times10^{-3}$, thus choosing $\ell_{\rm max}=6$ provides a high numerical accuracy.\footnote{Given the value of $\beta$, the quadruple precision is necessary to perform the computation at the desired order.}
Moreover, the associated Legendre polynomials are characterised by a symmetry property with respect to $\pi/2$, thus a suitable choice of the $N$ colatitudes $\theta_i$ allows us to divide the system in two subsystems, one for $\ell=0$ and even multipoles and the other for odd multipoles. The advantage of this separation is that of improving the accuracy of the solution because neglecting higher $\ell$'s produces an error which mainly arises only from the terms at $\ell_{\rm max}+2$ for even $\ell$ (or at $\ell_{\rm max}+1$ for odd $\ell$) \citep{2021A&A...646A..75T}.
The same consideration holds for different $\ell_{\rm max}$. 

Using the methods of elimination and substitution, the two subsystems can be easily solved. In particular, independently of the background spectral type, the $a_{\ell,m} (\nu, \beta)$ solution is given by a linear combination of sums and differences of the signals from Eq. \eqref{eq:eta_boost} evaluated at $\theta_{i}$ symmetrical with respect to $\pi/2$ \citep{2021A&A...646A..75T,2022ASPC..532..143B}.

For $\ell_{\rm max}=6$ and selecting $\theta_i = 0, \pi/4, \pi/3, \pi/2, (2/3)\pi, (3/4)\pi$, and $\pi$
(or the corresponding $w_i = {\rm cos} \, \theta_i = 1, \sqrt{2}/2, 1/2, 0, -1/2, -\sqrt{2}/2, -1$) in order to simplify the algebra, we have:
\begin{align}
\label{eq:struct_even}
a_{\ell,0} & = A_\ell \sqrt{\frac{4\pi}{2\ell+1}} \Bigg[ d_{\ell,1} \left({T_{\rm th}^{\rm BB/dist} (w=1) + T_{\rm th}^{\rm BB/dist} (w=-1) }\right) \nonumber
\\ & + d_{\ell,2} \left({T_{\rm th}^{\rm BB/dist} (w=\sqrt{2}/2) + T_{\rm th}^{\rm BB/dist} (w=-\sqrt{2}/2) }\right) \nonumber
\\ & + d_{\ell,3} \left({T_{\rm th}^{\rm BB/dist} (w=1/2) + T_{\rm th}^{\rm BB/dist} (w=-1/2) }\right) \nonumber
\\ & + d_{\ell,4} T_{\rm th}^{\rm BB/dist} (w=0) \Bigg] \,
\end{align}
\noindent
for $\ell=0$ and even multipoles, and 
\begin{align}
\label{eq:struct_odd}
a_{\ell,0} & = A_\ell \sqrt{\frac{4\pi}{2\ell+1}} \Bigg[ d_{\ell,1} \left({T_{\rm th}^{\rm BB/dist} (w=1) - T_{\rm th}^{\rm BB/dist} (w=-1) }\right) \nonumber
\\ & + d_{\ell,2} \left({T_{\rm th}^{\rm BB/dist} (w=\sqrt{2}/2) - T_{\rm th}^{\rm BB/dist} (w=-\sqrt{2}/2) }\right) \nonumber
\\ & + d_{\ell,3} \left({T_{\rm th}^{\rm BB/dist} (w=1/2) - T_{\rm th}^{\rm BB/dist} (w=-1/2) }\right) \Bigg]\,
\end{align}
\noindent
for odd multipoles. For the assumed $\theta_i$, Table \ref{tab:coeffs} gives the coefficients $A_\ell$ and $d_{\ell,i}$ \citep{2021A&A...646A..75T}. 

\begin{table}[h!]
    \caption{$A_\ell$ and $d_{\ell,i}$ coefficients.}
\begin{tabular}{cccccc}
\hline  \hline  \\ [-1.5ex]
$\ell$ & $A_\ell$ & $d_{\ell,1}$ & $d_{\ell,2}$ & $d_{\ell,3}$ & $d_{\ell,4}$ \\ [0.2ex]
\hline \\ [-1.ex]
0 & 1/630 & 29 & 120 & 64 & 204 \\[0.2ex]
1 & 1/210 & 29 & 60$\sqrt{2}$ & 32 & -- \\[0.2ex]
2 & 1/693 & 121 & 396 & $-$352 & $-$330 \\[0.2ex]
3 & 2/135 & 13 & 15$\sqrt{2}$ & $-$56 & -- \\[0.2ex]
4 & 8/385 & 9 & $-$10 & $-$16& 34\\[0.2ex]
5 & 32/189 & 1 & $-$3$\sqrt{2}$ & 4 & -- \\[0.2ex]
6 & 64/693 & 1 & $-$6 & 8 & $-$6 \\ [0.2ex]
\hline \hline & \\ [-1.5ex]
 \end{tabular}
    \label{tab:coeffs}
\end{table}

\subsection{Taylor's series expansion: harmonic coefficients and derivatives}
\label{sec:al0_deriv}

The solutions described by Eqs. \eqref{eq:struct_even} and \eqref{eq:struct_odd} have a structure that shows some similarities with the weights for the centred approximation numerical derivative scheme \citep{1988MC}, as already discussed in \cite{2021A&A...646A..75T}. This evidence indicates that there is a close relationship with the derivatives of the considered signal, as underlined for the first time in \cite{1981A&A....94L..33D} for the dipole.

We note that the values of $T_{\rm th}^{\rm BB/dist} (w)$ should be always very close to the one computed at $w=0$, i.e. in the direction perpendicular to the observer motion.
Let us assume that $T_{\rm th}^{\rm BB/dist} (w)$ can be expanded in Taylor's series around $w=0$.
Adopting the Lagrange notation, we denote with $T_{\rm th}^\dzero$, $T_{\rm th}'$, ..., $T_{\rm th}^\dsix$ the derivatives of
$T_{\rm th}^{\rm BB/dist} (w)$ performed with respect to $w$ evaluated at $w=0$, from order zero\footnote{Obviously, $T_{\rm th}^\dzero = T_{\rm th}$; we keep here the derivative index according to the notation in \cite{1988MC}.} to order six.
Thus, given the adopted set of $w_i$, after some calculations, Eqs. \eqref{eq:struct_even} and \eqref{eq:struct_odd} can be rewritten as:
\begin{align}
\label{eq:struct_der_even}
a_{\ell,0} & = 2\, A_\ell \sqrt{\frac{4\pi}{2\ell+1}} \Bigg[ \Bigg(d_{\ell,1} + d_{\ell,2} + d_{\ell,3}\Bigg)\, T_{\rm th}^\dzero \nonumber
\\ & + \frac{1}{2!} \,\Bigg(d_{\ell,1} + \frac{1}{2} d_{\ell,2} + \frac{1}{4} d_{\ell,3}\Bigg)\, T_{\rm th}^\dtwo \nonumber
\\ & + \frac{1}{4!} \,\Bigg(d_{\ell,1} + \frac{1}{4} d_{\ell,2} + \frac{1}{16} d_{\ell,3}\Bigg)\, T_{\rm th}^\dfour \nonumber
\\ & + \frac{1}{6!} \,\Bigg(d_{\ell,1} + \frac{1}{8} d_{\ell,2} + \frac{1}{64} d_{\ell,3}\Bigg)\, T_{\rm th}^\dsix + \frac{1}{2}\,d_{\ell,4}\, T_{\rm th}^\dzero \Bigg] \,
\end{align}
\noindent
for $\ell=0$ and even multipoles, and
\begin{align}
\label{eq:struct_der_odd}
a_{\ell,0} & = 2\, A_\ell \sqrt{\frac{4\pi}{2\ell+1}} \Bigg[ \Bigg(d_{\ell,1} + \frac{\sqrt{2}}{2}d_{\ell,2} + \frac{1}{2}d_{\ell,3}\Bigg)\, T_{\rm th}' \nonumber
\\ & + \frac{1}{3!} \,\Bigg(d_{\ell,1} + \frac{\sqrt{2}}{4}d_{\ell,2} + \frac{1}{8}d_{\ell,3}\Bigg)\, T_{\rm th}^\dthree \nonumber
\\ & + \frac{1}{5!} \,\Bigg(d_{\ell,1} + \frac{\sqrt{2}}{8}d_{\ell,2} + \frac{1}{32} d_{\ell,3}\Bigg)\, T_{\rm th}^\dfive \Bigg] \,
\end{align}
\noindent
for odd multipoles. Equations \eqref{eq:struct_der_even} and \eqref{eq:struct_der_odd} show that only the derivatives of even (odd) order contribute to $a_{\ell,0}$ for even (odd) $\ell$.
This is a consequence of the separation of the system into two subsystems, one for $\ell=0$ and even multipoles and the other for odd multipoles.

Finally, inserting the values of the $A_\ell$ and $d_{\ell,i}$ coefficients, after few algebra, we have:
\begin{equation}
\label{eq:struct_der_0}
a_{0,0} = \sqrt{4\pi} \; \left[T_{\rm th}^\dzero  + \frac{1}{6} T_{\rm th}^\dtwo + \frac{1}{120} T_{\rm th}^\dfour + \frac{1}{5040} T_{\rm th}^\dsix \right] \, ,
\end{equation}
\begin{equation}
\label{eq:struct_der_1}
a_{1,0} = \sqrt{\frac{4\pi}{3}} \; \left[T_{\rm th}'  + \frac{1}{10} T_{\rm th}^\dthree + \frac{1}{280} T_{\rm th}^\dfive \right] \, ,
\end{equation}
\begin{equation}
\label{eq:struct_der_2}
a_{2,0} = \frac{1}{3}\sqrt{\frac{4\pi}{5}} \; \left[T_{\rm th}^\dtwo + \frac{1}{14} T_{\rm th}^\dfour + \frac{1}{504} T_{\rm th}^\dsix \right] \, ,
\end{equation}
\begin{equation}
\label{eq:struct_der_3}
a_{3,0} = \frac{1}{15}\sqrt{\frac{4\pi}{7}} \; \left[T_{\rm th}^\dthree + \frac{1}{18} T_{\rm th}^\dfive \right] \, ,
\end{equation}
\begin{equation}
\label{eq:struct_der_4}
a_{4,0} = \frac{1}{105}\sqrt{\frac{4\pi}{9}} \; \left[T_{\rm th}^\dfour + \frac{1}{22} T_{\rm th}^\dsix \right] \, ,
\end{equation}
\begin{equation}
\label{eq:struct_der_5}
a_{5,0} = \frac{1}{945}\sqrt{\frac{4\pi}{11}} \; T_{\rm th}^\dfive \, ,
\end{equation}
\begin{equation}
\label{eq:struct_der_6}
a_{6,0} = \frac{1}{10395}\sqrt{\frac{4\pi}{13}} \; T_{\rm th}^\dsix\, ,
\end{equation}

\noindent where each denominator, $D_\ell$, in front of the square root can be rewritten as $D_\ell = (2\ell-1) D_{\ell-1}$, with $D_0 = 1$.
Equations \eqref{eq:struct_der_0}--\eqref{eq:struct_der_6} show that only the derivatives of order equal to or greater than $\ell$ contribute to $a_{\ell,0}$ and that the multiplicative factor in front of each derivative strongly decreases with the order of the derivative. We note that this property and the typical overall scaling of $a_{\ell,0} (\nu, \beta)$ almost proportional to $\beta^{\ell \cdot p}$ mentioned in Sect. \ref{sec:theoframe} do not imply that at each multipole $\ell$ the terms from the derivatives of order greater than $\ell$ are in general not relevant, since it is necessary to take into account the different frequency dependencies of the derivatives of different orders (we will delve deeper into this point after introducing background signals and other basic concepts, see Appendix \ref{appe}).

\subsection{Scaling of derivatives}
\label{sec:scal_deriv}

In the previous section we formally derived the link between the $a_{\ell,0}$ coefficients and the derivatives of
$T_{\rm th}^{\rm BB/dist} (w)$ with respect to $w$ evaluated at $w=0$. Here, we discuss the relationship between the derivatives performed with respect to $w$ and the ones performed with respect to the frequency, and their scaling relations relevant in next sections.

From Eq. \eqref{eq:eta_boost}, considering that $\nu' = \nu (1 - \beta w)/(1 - \beta^2)^{1/2}$ and omitting for simplicity the suffix `BB/dist', in the Leibniz notation we have
\begin{equation}\label{eq:linkder1}
\frac{dT_{\rm th}} {dw} = \frac{dT_{\rm th}} {d\nu'} \frac{d\nu'} {dw} = \frac{dT_{\rm th}} {d\nu'} \frac{-\beta \nu} {(1 - \beta^2)^{1/2}}\, 
\end{equation}
\noindent
for the first derivative, and
\begin{align}\label{eq:linkder2}
\frac{dT_{\rm th}^2} {dw^2} & = \frac{d} {dw} \left({ \frac{dT_{\rm th}} {dw} }\right) = 
\frac{-\beta \nu} {(1 - \beta^2)^{1/2}} \left[{ \frac{d} {d\nu'} \left({ \frac{dT_{\rm th}} {d\nu'} }\right) }\right] \frac{d\nu'} {dw} 
\\& = \frac{dT_{\rm th}^2} {d\nu'^2} \left[{ \frac{-\beta \nu} {(1 - \beta^2)^{1/2}} }\right]^2\, \nonumber
\end{align}
\noindent
for the second one.

In general, since the factor ${-\beta \nu} / {(1 - \beta^2)^{1/2}}$ does not contain $w$, for the subsequent $n$-th derivatives we have

\begin{equation}\label{eq:linkdern}
\frac{dT_{\rm th}^n} {dw^n} = \frac{dT_{\rm th}^n} {d\nu'^n} \left[{ \frac{-\beta \nu} {(1 - \beta^2)^{1/2}} }\right]^n\, .
\end{equation}

From Eqs. \eqref{eq:linkder1}--\eqref{eq:linkdern}, the derivatives $T_{\rm th}^\dn$ in Sect. \ref{sec:al0_deriv}, evaluated at $w=0$, can be rewritten setting  

\begin{equation}\label{eq:der_at_w0}
\frac{dT_{\rm th}^n} {d\nu'^n} \rightarrow {\frac{dT_{\rm th}^n} {d\nu'^n}}\Bigg|_{w=0} = {\frac{dT_{\rm th}^n} {d\nu'^n}}\Bigg|_{\nu'_{\beta,\perp}}\, ,
\end{equation}

\noindent where $\nu'_{\beta,\perp}$ is estimated at $w=0$ (or $\theta=\pi/2$)

\begin{equation}\label{eq:nuperp}
\nu'_{\beta,\perp} = \frac{\nu} {(1 - \beta^2)^{1/2}} \, .
\end{equation}

Considering a speed, $\beta_a$, different from $\beta$, the ratio between the derivatives $T_{\rm th}^\dn$ computed for these two different speeds is
\begin{equation}\label{eq:ratio_der}
\frac{{T_{\rm th}^\dn}\Bigr|_{\beta}} {{T_{\rm th}^\dn}\Bigr|_{\beta_a}}  =  f_a^{-n} \, \left( { \frac{1-\beta_a^2}{1-\beta^2} } \right)^{n/2} \, R_n \, ,
\end{equation}

\noindent where $f_a = \beta_a /\beta$ and 

\begin{equation}\label{eq:R}
 R_n =  \Bigg ( {\frac{dT_{\rm th}^n} {d\nu'^n}}\Bigg|_{\nu'_{\beta,\perp}} \Bigg ) \, \Bigg ( {\frac{dT_{\rm th}^n} {d\nu'^n}}\Bigg|_{\nu'_{\beta_a,\perp}} \Bigg )^{-1}
\end{equation}

\noindent being $\nu'_{\beta_a,\perp}$ defined as in Eq. \eqref{eq:nuperp} but for $\beta = \beta_a$.

For $\beta$ and $\beta_a$ significantly less than unit $\nu'_{\beta_a,\perp} \simeq \nu'_{\beta,\perp}$.
All the three factors in the right hand side of Eq. \eqref{eq:ratio_der} are in principle different from unit. On the other hand, except for possible functions $T_{\rm th}$ with extreme variation in frequency, $ f_a^{-n}$ is the only term that can be significantly different from unit.
We note that the extremely accurate computation of $R_n$, i.e. of its very little difference from unit, would require an analogous knowledge of the change of the corresponding order derivative in an extremely narrow range between ${\nu'_{\beta,\perp}}$ and ${\nu'_{\beta_a,\perp}}$, which obviously is the missing information in the problem under consideration, calling for the methods studied in this work. In practice, we will set $R_n$ = 1. Of course, this limitation, which treatment is out of the scope of this work, is much less critical than the instability problem object of this work.\footnote{In principle, even in the absence of a good analytic or semi-analytic representation of the monopole spectrum on its whole frequency range, it could be possible to estimate $R_n$ for different analytic representations among the more reasonable ones in a given frequency range of particular interest.}

\section{Monopole spectrum tabulated models}
\label{sec:monmod}

\subsection{ESMB -- smooth spectral shape}
\label{sec:microw_back}

Extragalactic radio and microwave sky, from hundreds of MHz to few hundreds of GHz, is dominated by sources powered by AGN, in which the observed flux density is produced by synchrotron radiation generated by the acceleration of relativistic charged particles. The frequency spectra of radio AGN, generally characterised by a power law ($S\propto\nu^{\alpha}$) with a steep ($\alpha<-0.5$) or flat ($-0.5<\alpha<0.5$) slope depending on observed fluxes, mainly originate in extended (optically thin) radio lobes or in compact (optically thick) regions of the radio jet, respectively. 
In the \lq unified model\rq\, \citep[e.g.,][]{urr95,net15}, these two populations arise from the different orientation of the observer relative to the axis of the characteristic jets emerging from the central black hole: in the case of a side-on view of the jet-axis, the observed (low-frequency) emission emerges from the extended lobes, with a typical steep spectrum.
On the other hand, if the line of sight is close to the axis of the emitting jet, objects appear as compact flat-spectrum sources, and are referred to as blazars \citep[e.g.,][]{dez10}. Due to orientation effects, steep-spectrum sources are much more numerous than blazars and, thus, are the most relevant population at classical radio frequencies, below $10-20$ GHz. However, because of the spectral behaviour, their relevance reduces increasingly with frequency and they become sub-dominant with respect to flat-spectrum sources starting from few tens of GHz.

Number counts of extragalactic radio sources are well determined at radio frequencies $\nu\lsim10$\,GHz down to flux densities of $S\ll1$\,mJy thanks to deep and large-area surveys \citep[e.g..][]{dez10,bon11,mas11,2012ApJ...758...23C, mil13,smo17, huy20}. Luminosity functions and multi-frequency number counts of radio sources are also well modelled at these frequencies \citep[e.g.,][]{tof98,dez05,mas10,tuc11,tuc21}. Very recently, \citet{2023MNRAS.521..332T} have published a thorough discussion of source number counts at centimetre wavelengths showing that the AGN population is dominating the number counts down to flux densities $S\gsim 1$\,mJy. At higher frequencies, i.e. from tens of GHz to millimetre wavelengths, observational data on radio sources are mainly provided by CMB experiments \citep[e.g.,][]{pla16xxvi,dat19,gra20,eve20}, which are able to detect only bright sources, down to tens of mJy at best. The uncertainties on number counts are still large, especially in the frequency range where the CMB dominates, that is, between 70 and 300 GHz.

In this analysis, we use the differential number counts, $n_{\nu}(S)$, of extragalactic radio sources at radio/microwave frequencies provided by the \citet{tuc11} model, in its updated version \citep{lag20} based on recent data from the Atacama Cosmology Telescope (ACT) \citep{dat19} and South Pole Telescope (SPT) \citep{eve20} experiments. Similarly to the evolutionary model of \citet{dez05}, radio sources in the NRAO VLA Sky Survey (NVSS) and Green Bank 6-cm (GB6) surveys \citep{gre96,con98} are separated in steep- and flat-spectrum sources according to their spectral behaviour measured between 1 and 5\,GHz. Flux densities of radio sources are then extrapolated to higher frequencies by considering characteristics of the physical mechanisms of emission \citep{bla79,kon81} for the different source populations identified (differently from \citet{dez05}, who apply a simple power law extrapolation of radio spectra to higher frequencies). In particular, in the \citet{tuc11} model, the spectrum of flat-spectrum sources is expected to break at some frequency in the range of $10-1000$ GHz and to steepen at higher frequencies, due to electrons cooling effects and to the transition of the observed synchrotron emission from the optically thick to the optically thin regime. The frequency break is different for flat-spectrum radio quasars (FSRQs) and for BL\,Lac objects. In the former population, the spectral break is expected to typically occur at $\nu<100$\,GHz, while in BL\,Lacs it should appear at $\nu\gsim100$\,GHz (implying more compact emitting regions than FSRQs). The \citet{tuc11} model,\footnote{We always refer here to the most successful model discussed in that paper and called \lq C2Ex\rq. See \citet{tuc11}, section 4, for more details.} and its update published in \cite{lag20}, provides a good fit of observational number counts from all the CMB experiments at frequencies between 30 and 220\,GHz, down to flux densities of 10\,mJy or less. Moreover, as discussed by \citet{mas22}, the \lq C2Ex\rq model is able to give a very good fit also to the very recent number counts calculated from a complete sample of blazars selected by the {\it Herschel} Astrophysical Terahertz Large Area Survey (H-ATLAS). The model appears also in good agreement with the \lq extragalactic radio background light\rq\, at GHz frequencies estimated by \citet{2023MNRAS.521..332T}, as there defined, calculated without excluding the brightest discrete extragalactic sources, i.e. without a high flux density limit: the model finds 105, 10.6, 2.4, and 0.57\,mK at 1.4, 3, 5, and 8.4\,GHz, respectively, very close to the values reported by \citet{2023MNRAS.521..332T} in their table 6B, third column, \lq\, AGN-fit EBL\rq.

\begin{figure}[h!]
\centering
         \includegraphics[width=9.cm]{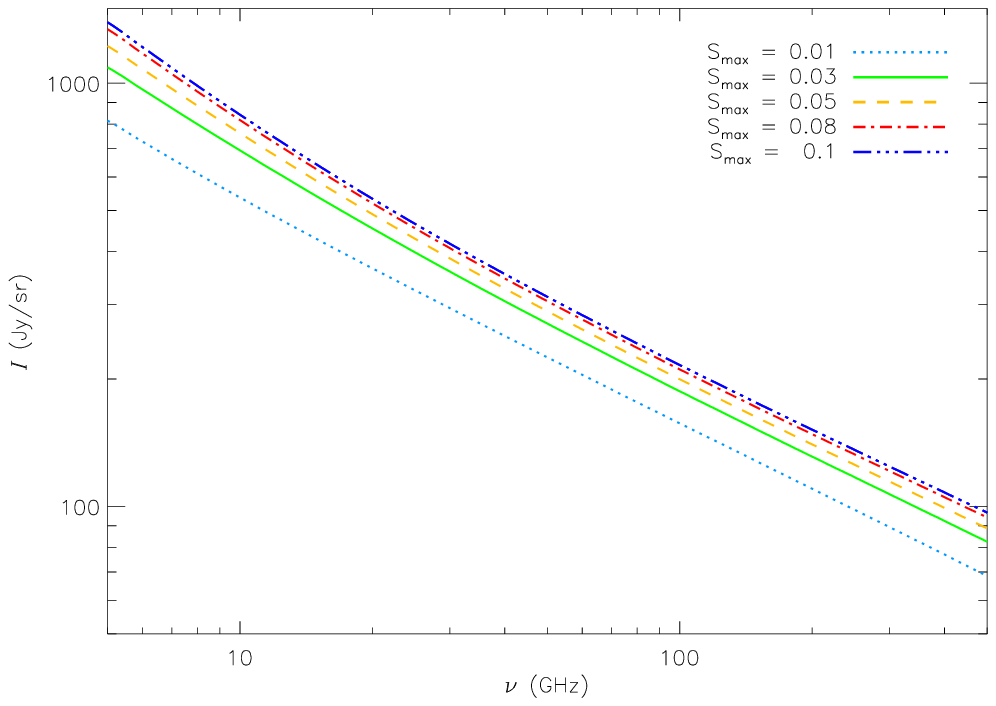}
    \caption{Comparison among the tabulated intensities of the considered ESMB model (C2Ex) for the different detection thresholds $S_{max}$ discussed in the text. See legend.}
    \label{fig:Back_mm_Int}
\end{figure}

Based on the modelled differential number counts we estimate the extragalactic background intensity between 1\,GHz and 1\,THz,\footnote{In the rest of the work, we will show our results between 5\,GHz and 500\,GHz (see Fig.\,\ref{fig:Back_mm_Int}) where the model is better probed and not overwhelmed by other extragalactic background contributions, particularly at higher frequencies. Furthermore, working in a restricted frequency range allows us to be safe with respect to potential boundary effects.} in a grid of 512 frequencies equispaced in logarithmic units. The background intensity, $I(\nu) = (2 h / c^2)\, \nu^3\, \eta(\nu) = (2 k / c^2)\, \nu^2 \, T_{\rm ant}(\nu)$, being $T_{\rm ant} (\nu) = (h\nu/k)\,\eta(\nu)$ the corresponding antenna temperature, is proportional to $\int_{S_{min}}^{S_{max}}\,S\,n_{\nu}(S)\,dS$, where $S_{max}$ corresponds to the detection threshold above which sources are detected and removed from data, while $S_{min}$ is the minimum flux density considered (sources with flux densities below this value should add a negligible contribution to the above integral). The background intensity is shown in Fig.\,\ref{fig:Back_mm_Int} for five values of $S_{max}$, from 0.01 to 0.1\,Jy. The largest value corresponds (approximately) to the detection limit of all-sky experiments like {\it Planck}, while the smallest one to the limit of high-resolution CMB experiments like SPT and ACT. On the other hand, number counts are modelled down to $S_{min}\sim10\,\mu$Jy. Below this value, the contribution of classical extragalactic radio sources (i.e. mainly steep-spectrum giant ellipticals and quasars (QSOs)) is not expected to be completely negligible. For example, \citet{tuc21} modelled luminosity functions of radio-loud AGN at GHz frequencies and provided estimates of number counts at $1-15$\,GHz down to $1\,\mu$Jy. Based on those results, we find that changing $S_{min}$ from 10 to $1\,\mu$Jy increases the background intensity estimates by a few per cent, in the range of the considered $S_{max}$. Although the missing contribution can be larger at microwave wavelengths, our estimates of the background intensity should be a good approximation of the true one, and the choice of minimum flux $S_{min}=10\,\mu$Jy should not affect the conclusions of the current analysis. From Fig.\,\ref{fig:Back_mm_Int}, it can be appreciated the change in slope of the extragalactic background at $10-50$\,GHz: at higher frequencies the frequency spectrum flattens due to the fact that blazars become more and more dominant over steep-spectrum extragalactic sources.

\begin{figure}[h!]
\centering
         \includegraphics[width=9.cm]{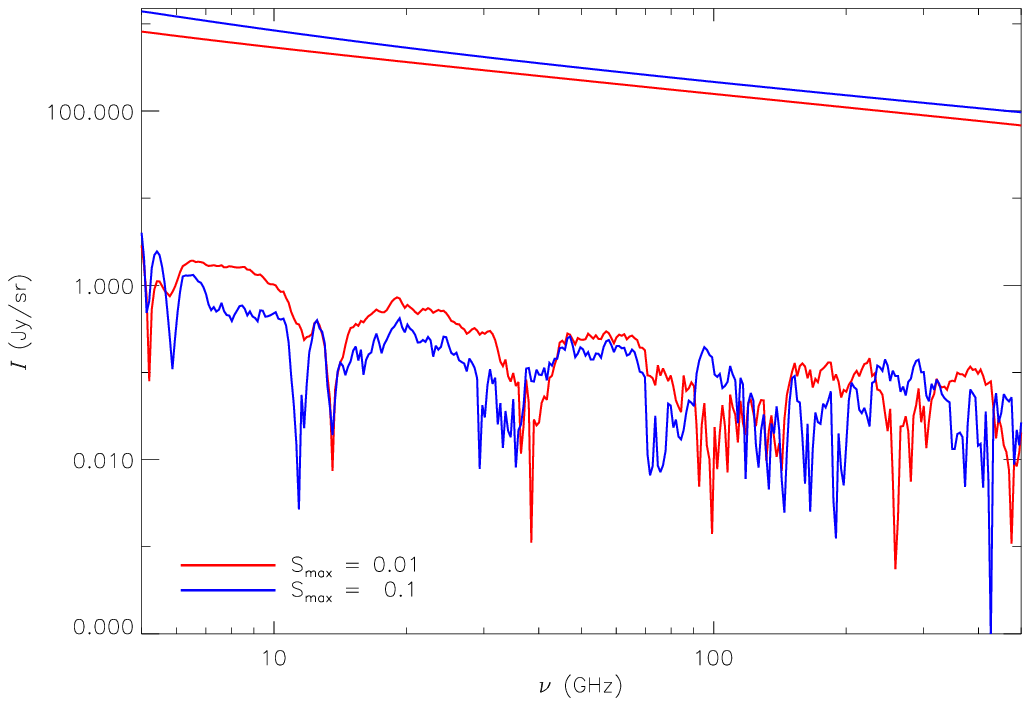}
    \caption{Absolute value of the difference between the log-log polynomial interpolations and the original tabulations (bottom curves) for the maximum and minimum detection thresholds adopted in the calculation of the ESMB total intensity (top curves). See legend.}
    \label{fig:Back_mm_res}
\end{figure}

In order to have a reference analytical representation of the ESMB intensity, we fitted log\,$I(\nu)$ as function of log\,$\nu$ with polynomials of different degrees.\footnote{In passing, we note that this type of analytical function belongs to the set of Derivative Constrained Functions discussed in \cite{2021MNRAS.502.4405B} and it has been adopted in e.g., \cite{2021APh...12602532N} to represent the cosmic radio background spectrum.} 
For these five detection thresholds, we found that a nine degree polynomial, with coefficients appropriate to each case, fits very well the tabulated ESMB intensity: as shown in Fig. \ref{fig:Back_mm_res} for the maximum and minimum values of $S_{max}$, the relative difference between the fitting polynomial and the original tabulation turns to be between $\sim 10^{-4}$ and $\sim 10^{-3}$.
This range of values can be assumed as a reasonable estimate of the relative uncertainty of the functional representation of the ESMB. The statistical errors due to the different sample sizes of extragalactic sources that are used to estimate their number counts are, instead, much larger (typically $\gsim 5-10$\%).\footnote{A direct consequence of the limited number of sources present in each sample, is that the division of the observed data set in a convenient number of flux density bins introduces statistical fluctuations in the calculations of source number counts.}
 
\subsection{Redshifted 21cm line -- feature-rich spectral shape}
\label{sec:21cm}

The 21cm line corresponds to the spin-flip transition in the ground state of neutral hydrogen (HI). This signal is described as the offset of the 21cm brightness (i.e. antenna) temperature from the background temperature, $T_{\rm back}$, along the observed line of sight at a frequency $\nu$ that, because of cosmic expansion, is related to the rest frame frequency, $\nu_{21 {\rm cm}} = c / (21 {\rm cm})$, by $\nu = \nu_{21 {\rm cm}} / (1+z)$. $T_{\rm back}$ is usually assumed equal to $T_r$ but, in general, it could include potential distortions and other radiation backgrounds. Since the signal detected at a given frequency corresponds to a specific redshift, the 21cm line provides a tomographic view of the cosmic evolution. 

A rich set of redshifted 21cm line models has been presented in \cite{2017MNRAS.472.1915C}, resulting in a wide envelope of predictions for the antenna temperature.
The publicly available 21cm Global EMulator code\footnote{\url{https://people.ast.cam.ac.uk/~afialkov/index.html}} (21cmGEM) predicts, from a large dataset of standard astrophysical models, the global 21cm signal over the redshift interval between 5 and 50 including both the epoch of reionization (EoR) and Cosmic Dawn. Each model corresponds to a specific row number ($\#_{r}$) in the dataset and is based on seven key astrophysical parameters: the star formation efficiency ($f_{*}$), the minimum virial circular velocity of star-forming haloes ($V_{c}$), the X-ray radiation efficiency ($f_{X}$), the CMB optical depth ($\tau$), the slope ($\alpha$) and minimum energy ($\upsilon_{min}$) of the X-rays spectral energy distribution (SED) and the ionizing efficiency of sources ($R_{\rm mfp}$) \citep{10.1093/mnras/staa1530}. From a total of 2186 test models computed with the 21cmGEM code, we extract six scenarios that cover a sufficiently wide range of different shapes and signals. The names, row numbers and key parameters of these models are reported in Table \ref{tab:21param}.

Detecting and characterizing the redshifted 21cm line from the diffuse HI in the intergalactic medium is experimentally very challenging and requires an extremely accurate subtraction of the much more intense foreground signals.
A pronounced absorption profile centred at $(78 \pm 1)$ MHz has been found by \cite{EDGESobs2018Nature} based on the data from the Experiment to Detect the Global EoR Signature (EDGES).
The authors described the absorption profile in terms of a flattened Gaussian identified by a set of best-fit parameters.\footnote{Some works alternatively explain the EDGES profile in terms of residual instrumental systematic effects \citep[e.g.,][]{2020MNRAS.492...22S} or considered the impact of ionosphere \citep{shen_etal_2021}, while the results of the Shaped Antenna measurement of the background RAdio Spectrum (SARAS) experiment \citep{2022NatAs...6..607S} are not in agreement with the EDGES cosmological signal profile.} This representation, being fully analytical, is adopted as reference model to test the quality of the methods studied in this paper.

\begin{table*}[h!]
    \caption{Model name, its corresponding row number and the seven key parameters for the 21cm signals from the 21cmGEM code.}
\begin{tabular}{ccccccccc}
\hline  \hline  \\ [-1.5ex]
Model & $\#_{r}$ & $f_{*}$ & $V_{c}$ (km/s) & $f_{X}$ & $\tau/10^{-2}$ & $\alpha$ & $\upsilon_{min}$ (keV) & $R_{\rm mfp}$ (Mpc) \\ [0.3ex]
\hline \\ [-1.ex]
A & 821 & 5 $\cdot\, 10^{-3}$ & 76.5 & $10^{-3}$ & 7.8120867 & 1 & 0.3 & 20 \\[0.2ex]
B & 1494 & 0.5 & 4.2 & $10^{-3}$ & 6.2411525 & 1.3 & 0.1 & 40 \\[0.2ex]
C & 1394 & 1.2584272 $\cdot\, 10^{-2}$ & 8.5060319 & 4.6333723 & 7.5685575 & 1.5 & 3 & 49 \\[0.2ex]
D & 1291 & 0.17854233 & 66.365130 & 4.2258934 & 6.5817279 & 1.3 & 3 & 11 \\[0.2ex]
E & 376 & 0.5 & 24.2 & $10^{-5}$ & 6.6450808 & 1.5 & 0.2 & 35 \\[0.2ex]
F & 2168 & 6.9054805 $\cdot\, 10^{-3}$ & 32.431833 & 0.22365493 & 4.7274618 & 1 & 0.1 & 12.976633 \\[0.2ex]
\hline \hline & \\ [-1.5ex]
 \end{tabular}
    \label{tab:21param}
\end{table*} 

\begin{figure}[h!]
\centering
         \includegraphics[width=9.cm]{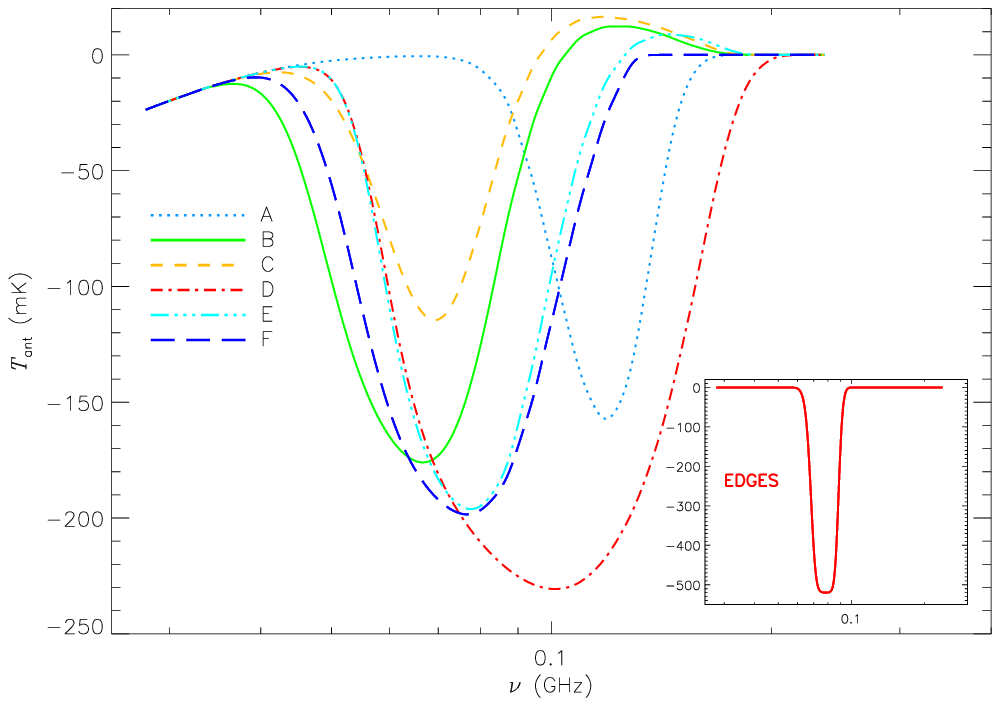}
    \caption{Comparison between the considered redshifted 21cm line models. See legend and text.}
    \label{fig:21cmModels}
\end{figure}

In Fig. \ref{fig:21cmModels}, we show the extracted simulated models and the anomalously strong and narrow absorption feature of EDGES, the latter displayed in the inset since its more pronounced temperature decrement with respect to the others.

The 21cm signals are produced in a redshift equispaced grid of 451 points with a 0.1 step from $z=50$ to $z=5$, i.e. in a frequency range between $\sim 27.85$ GHz and $\sim 236.73$ GHz. The B, C and E models are characterised by a temperature decrement that, above a certain frequency, becomes positive, differently from the other scenarios and from the EDGES profile. We adopt the same grid to also produce the tabulation based on the EDGES analytical representation. 

\section{Instability of harmonic coefficients with direct formulas}
\label{sec:instability}

We present here the results of the computation of the transfer of the monopole frequency spectrum to the higher multipoles 
applying directly the analytical solutions given in Sects. \ref{sec:sol_struct}--\ref{sec:scal_deriv}.

As in \cite{2021A&A...646A..75T}, we show our results in terms of spherical harmonic coefficients $a_{\ell,0} (\nu,\beta)$ from $\ell = 1$ to $\ell_{\rm max} = 6$ but
expressed in terms of their difference, $\Delta a_{\ell,0}$, with the coefficients $a_{\ell,0} (\nu,\beta)^{\rm BB}$ obtained in the case of a monopole frequency spectrum represented by a blackbody at the temperature $T_0$.
Particularly for low signals, the relevant information is indeed contained in the differences of the $a_{\ell,0} (\nu,\beta)$ coefficients with respect to the ones corresponding to a suitable reference case. 
Typically, the observer velocity is assumed to be the same with respect to a frame at rest with the CMB or other extragalactic backgrounds. The scaling with $\beta$ (see Sect. \ref{sec:scal_deriv}) of the $a_{\ell,0} (\nu,\beta)$ coefficients for the CMB and the backgrounds is the same, thus $\Delta a_{\ell,0}$ scale accordingly.\footnote{In general, the $a_{\ell,0} (\nu,\beta)$ should be computed separately, and, for different velocity directions, one can assume the $z$-axis parallel to each velocity and then rotate \citep{1984JGR....89.4413G} the $a_{\ell,0} (\nu,\beta)$ to a common reference system, implying that $a_{\ell,m} (\nu,\beta)$ with $m \ne 0$ do not vanish.} 

For the redshifted 21cm line, as a reference case in the calculation of the difference $\Delta a_{\ell,0}$, one could also consider, for comparison, the combined signal of a BB at the temperature $T_0$ plus a suitable representation of the radio extragalactic background spectrum, that indeed dominates over the standard CMB BB spectrum at $\nu \lsim 1$ GHz (see e.g., appendix A in \cite{2020PhRvR...2a3210B} for a recent data compilation).

Although we mainly focus at $\ell \ge 1$, in the next sections we will consider also the results about the ratio, $R = (a_{0,0} (\nu,\beta) / \sqrt{4\pi}) / T_{\rm th} (\nu)$, 
between the equivalent thermodynamic temperature of observed and intrinsic monopole, expressing it in terms of the difference \citep{2021A&A...646A..75T} $\Delta R = R - R^{\rm BB}$, where, theoretically, 
$R^{\rm BB} \simeq (1 - 2.5362 \times 10^{-7})$ is the same ratio but for the case of the blackbody at the present temperature $T_0$. Indeed, as explicitly shown by Eq. \eqref{eq:struct_der_0}, the contribution from high order derivatives appears also in $a_{0,0}$.

\subsection{Interpolation versus derivatives}
\label{sec:InterpVSderiv}

For a monopole frequency spectrum described by a tabulated function, the evaluation of the quantities $T_{\rm th}^{\rm BB/dist} (w)$ or $T_{\rm th}^\dn$ requires the interpolation of the corresponding functions in the adopted grid of points. \cite{1988MC,1998SIAMR..40..685F} provided the weights for the (possibly centred) approximations at a grid point for the generation of finite difference formulas on arbitrarily spaced grids for any order of derivative. We are interested in evaluating derivatives from order zero up to order six. For a fixed number of grid points to be used in the differentiation scheme, $n_{\rm ds}$, the order of accuracy decreases with the order of derivative. We performed some tests increasing $n_{\rm ds}$ or fixing the order of accuracy and working with variable $n_{\rm ds}$ according to the order of derivative, without finding significant differences provided that at least the fourth order of accuracy at the highest derivative is achieved. For simplicity, we work with a fixed $n_{\rm ds} = 9$ achieving the fourth order of accuracy up to the derivative of order six. For each value of $\nu'_{\beta,\perp}$, i.e. $w=0$, (typically, for all the {\it inner} points of the grid), we select the grid point closest to $\nu'_{\beta,\perp}$ and 4 points on the left and 4 on the right, to work with an almost centred approximation.\footnote{If the desired value of $\nu'_{\beta,\perp}$ is close to the first left (right) boundary point of the grid, we could consistently select more grid points on the right (left), resulting e.g., into an almost one-sided approximation. In practice, to fully prevent possible effects associated to the transition between the two types of approximations, we prefer to avoid the computation for the $(n_{\rm ds} -1)/2$ points closer to each of the grid extremes.}  We can then directly compute the $a_{\ell,0}$ coefficients using the weights in \cite{1988MC} for the zero order derivative in the desired value of $w$ when using the equations in Sect. \ref{sec:sol_struct} (we will call this {\it the interpolation scheme}), or the ones for the derivative from order zero up to order six but at $w=0$ when using the equations in Sects. \ref{sec:al0_deriv}--\ref{sec:scal_deriv} (we will call this {\it the derivative scheme}). 

We add, in terms of antenna temperature, the background and the CMB BB monopole spectra and compute the corresponding $a_{\ell,0}$ coefficients\footnote{Indeed, the types of background considered here are superimposed to the CMB. For simplicity, we consider here a Planckian law at temperature $T_0$ for the CMB spectrum, although it could be replaced by a distorted CMB spectrum, as discussed in \cite{2021A&A...646A..75T} for analytic or semi-analytic functions.} using the interpolation and derivative schemes.
In general, the $a_{\ell,0} (\nu,\beta)^{\rm BB}$ terms entering in the differences $\Delta a_{\ell,0}$ and $\Delta R$ are computed with the same treatment.

To focus on the issues raised applying directly the analytical solutions in Sects. \ref{sec:sol_struct}--\ref{sec:scal_deriv}, we consider the reference cases based on the log-log polynomial characterisation of the ESMB and the analytical representation of the EDGES absorption profile (Sect. \ref{sec:monmod}). We compute the monopole spectrum at each grid frequency and pollute it with a simulated noise to mimic potential inaccuracies affecting the functional representation of the background. These inaccuracies, and not the much larger real astrophysical model uncertainty, are indeed the types of errors that amplify at the increasing derivative order. They are modelled as the sum of two terms added to the signal in terms of $T_{\rm ant}$ to give

\begin{equation}\label{eq:T&N}
T_{\rm ant}^{\rm ESMB/21cm} = T_{\rm ant, analytical}^{\rm ESMB/21cm} \, \left({1 +  r_{\rm err}\, G_1}\right) + a_{\rm err}\, G_2 \, ,
\end{equation}

\noindent
where $G_1$ and $G_2$ are extracted from Gaussian (pseudo)random realisations with zero average and unit variance, the dimensionless constant $r_{\rm err}$ determines the amplitude of the inaccuracies proportional to the signal and the constant $a_{\rm err}$ gives the amplitude of the inaccuracies independent of the signal.
Of course, setting $r_{\rm err} = a_{\rm err} = 0$ one can study the effect coming just from the discretisation associated to the adopted grid and from the adopted scheme.
The results obtained in both the schemes can be then easily compared with those derived in the ideal fully analytical case.

The original tabular representations of the background monopole spectra already include their own potential inaccuracies. Thus, we avoid to add any further simulated inaccuracy.

\subsection{ESMB models}
\label{sec:probl_microw_back}

\begin{figure*}[h!]
\hskip -0.6 cm
         \includegraphics[width=10cm]{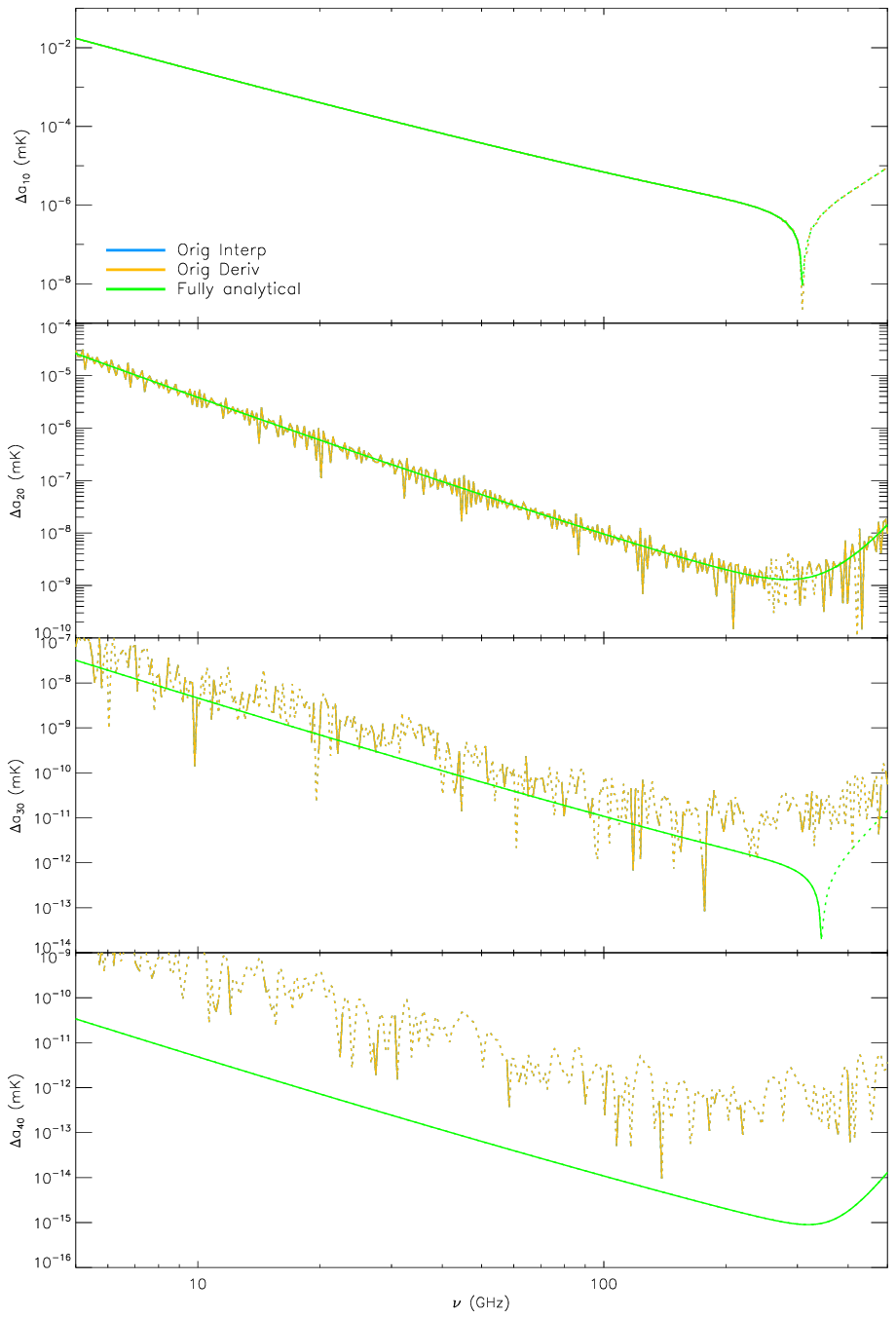}
\hskip -0.7 cm
         \includegraphics[width=10cm]{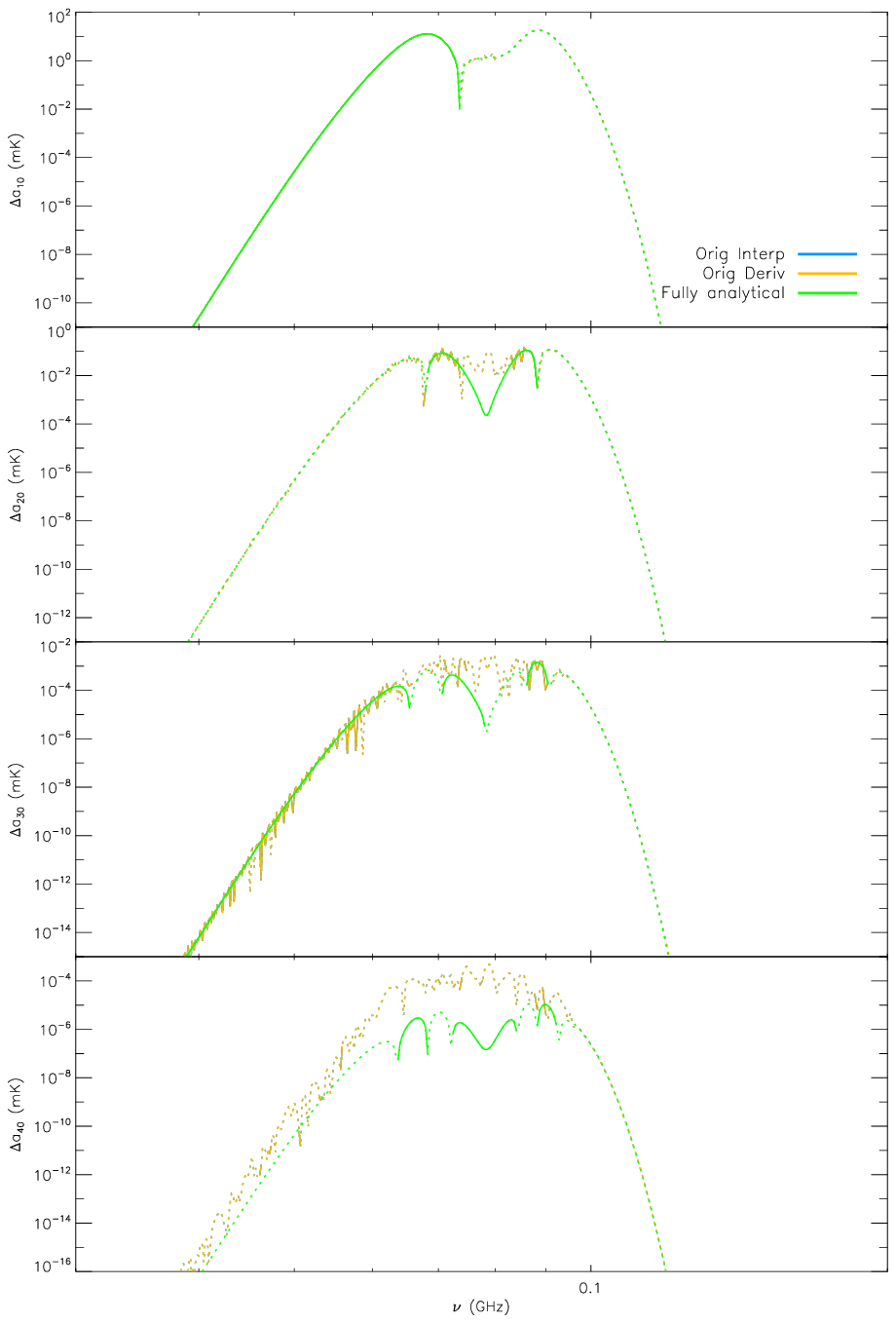}
    \caption{Comparison between the interpolation and the derivative schemes with the ideal fully analytical treatment applied to the analytical representations of the ESMB (left column) and the 21cm EDGES profile (right column) considering $r_{\rm err}^{{\rm ESMB}} = 2.5 \times 10^{-4}$ and $r_{\rm err}^{21{\rm cm}} = 10^{-3}$, respectively, without including the pre-filtering and up to multipole 4. Here (and in the following analogous figures), solid (dots) lines refer to positive (negative) values. See legend.}
    \label{fig:original}
\end{figure*}

We apply the method described above to the ESMB discussed in Sect. \ref{sec:microw_back}. Assuming the polynomial representation to characterise the ESMB, Fig. \ref{fig:Back_mm_res} shows that the inaccuracies in the tabulated ESMB are almost proportional to the ESMB intensity and we then set $a_{\rm err} = 0$. For conciseness, we report the case of the largest threshold $S_{max} =0.1$\,Jy, adopting a value $r_{\rm err} = 2.5 \times 10^{-4}$, an \lq average\rq\, value between the maximum and minimum relative differences displayed in Fig. \ref{fig:Back_mm_res}. The results at $1 \le \ell \le 4$ are shown in the left panels of Fig. \ref{fig:original}. As expected, we also find that for the choice $r_{\rm err} = a_{\rm err} = 0$ the agreement between the results based on both the interpolation and the derivative schemes and on the ideal fully analytical calculation is extremely good, 
and we then avoid, for simplicity, to display this case in the figure. 
We also note that including simulated noise the interpolation and the derivative schemes give very similar results. 
Remarkably, even for $r_{\rm err} = 2.5 \times 10^{-4}$, numerical uncertainties in the ESMB monopole spectrum propagate to higher $\ell$, as evident from the figure already at $\ell = 2$, and their effect dramatically increases with $\ell$, making the $\Delta a_{\ell,0}$ computation highly unstable. The effect is particularly evident where $\Delta a_{\ell,0}$ should present the change of sign and, in general, at low values of $\Delta a_{\ell,0}$.

The same analysis applied to the original tabulation, reported in Appendix \ref{appa} for simplicity, gives essentially the same result indicating that the above issue is indeed due to intrinsic uncertainties present in the calculation of the ESMB spectrum. Also, the similarity between the amplitude of the effect found in the two cases suggests that $r_{\rm err} = 2.5 \times 10^{-4}$ represents a reasonable choice to globally characterise the inaccuracies in the ESMB tabulated spectrum for $S_{max} =0.1$\,Jy.

\subsection{Redshifted 21cm line models}
\label{sec:probl_21cm}

We perform the same kind of analysis for the redshifted 21cm line. To estimate representative values of $r_{\rm err}$ and $a_{\rm err}$, we consider the results discussed in \cite{2021MNRAS.508.2923B} about the overall uncertainty of numerical codes for the prediction of the global 21cm signal. According to the authors, their current precision is of about a few per cent or, in absolute sense, around the mK level. We consider here values of $r_{\rm err}$ and $a_{\rm err}$ around an order of magnitude less, being interested in the uncertainty within relatively narrow frequency ranges, comparable to those adopted in the implementation of the differentiation scheme that covers about ten points of the grid, and also in view of the great efforts in this field expected for the future. 
Thus, we first consider the analytical representation of the EDGES absorption profile and adopt $r_{\rm err} = 10^{-3}$. The results at $1 \le \ell \le 4$ are shown in the right panels of Fig. \ref{fig:original}. Also in this case, for $r_{\rm err} = a_{\rm err} = 0$ we verified the extremely good agreement between both the interpolation and the derivative schemes and the ideal fully analytical calculation, as well as that, in general, the interpolation and the derivative schemes give very similar results. 
As found in the case of the ESMB, numerical uncertainties in the redshifted 21cm line monopole spectrum clearly propagate to higher $\ell$ and their effect significantly increases with $\ell$, being particularly evident at low values of $\Delta a_{\ell,0}$ and around frequencies corresponding to the sign changes whose occurrence increases with $\ell$.

With respect to the case $r_{\rm err} > 0$ and $a_{\rm err} = 0$, for $r_{\rm err} = 0$ and $a_{\rm err} > 0$ the effect of numerical uncertainties is relatively smaller (larger) at larger (smaller) absolute values of $\Delta a_{\ell,0}$ (Appendix \ref{appb}). 

In general, we find that for larger numerical uncertainties the error propagation effect appears even at $\ell = 1$, as shown for the ESMB and the EDGES absorption profile respectively in Appendix \ref{appc} and \ref{appd}, where we report on the instability mitigation achieved with the methods presented in the following section considering, for instance, values of $r_{\rm err}$ one order of magnitude larger or smaller than the reference values adopted above.

Finally, we tested the impact of a different assumption for the reference background in the calculation of the $\Delta a_{\ell,0}$. For example, we considered an overall radio background at low frequency represented by equation 5, section 3, of \cite{2018ApJ...858L...9D}. As expected, the results expressed in terms of $\Delta a_{\ell,0}$ do not change with respect to those reported in Fig. \ref{fig:original}.

\section{Filtering}
\label{sec:filt}

In the previous section we have shown that inaccuracies in the tabulated monopole frequency spectrum can prevent a robust and accurate computation of its transfer to higher multipoles.
To solve this problem, we describe a pre-filtering method (Sect. \ref{sec:prefilt}) which can be used independently or in combination with the subsequent filtering techniques (Sects. \ref{sec:filt_der} and \ref{sec:filt_amp_deamp}) which, in turn, can also be applied independently.

\subsection{Pre-filtering of monopole spectrum}
\label{sec:prefilt}

A first approach is based on the pre-filtering of the tabulated monopole frequency spectrum in order to smooth out inaccuracies occurring at small scales in a suitable equispaced real space variable, $u$. 
We consider a low-pass Gaussian filtering in the Fourier space: we (i) first compute\footnote{\url{http://www.phys.ufl.edu/~sazonov/fft.f}} the fast Fourier transform (FFT) of the monopole spectrum tabulated in a grid $u_i$ ($i=1,N_p$, with $N_p$ a power of 2), then (ii) smooth it out at the modes, $f_i = 1, N_p$, corresponding to the small scales in the real space, and finally (iii) obtain a filtered monopole spectrum applying the inverse FFT (FFT$^{-1}$) to the smoothed FFT, namely:
\begin{align}\label{eq:prefilt}
{\rm (i)} \;\; & T_{\rm th}^{\rm BB/dist} (u_i) \rightarrow F(f_i) = {\rm FFT} (T_{\rm th}^{\rm BB/dist} (u_i)) \nonumber
\\ {\rm (ii)} \;\; & F(f_i) \rightarrow F_s(f_i) = F(f_i)  [{\rm e}^{-(f_i/\sigma_f)^2} + {\rm e}^{-((f_i-N_p)/\sigma_f)^2} ]
\\ {\rm (iii)} \;\; & F_s(f_i)  \rightarrow T_{\rm th,s}^{\rm BB/dist} (u_i)=  Real \left({ {\rm FFT}^{-1} (F_s(f_i)) }\right)\, , \nonumber
\end{align}
\noindent
where in (iii) we obviously take the real part since $T_{\rm th}^{\rm BB/dist}$ is a real function. In Eq. \eqref{eq:prefilt}, $\sigma_f  = f N_p$, where $f$ determines the level of smoothing, which decreases for increasing $f$, in terms of fraction of modes. We tested the effect of different values of $f$ in a range from $\simeq 0.05$ to $\simeq 0.2$, larger (smaller) values typically resulting into a negligible smoothing (into an excessive smoothing, possibly significantly affecting the original background shape). The filter in Eq. \eqref{eq:prefilt} is symmetric around the central mode, according to the convention adopted for the FFT. Given the tabulations described in Sects. \ref{sec:microw_back} and \ref{sec:21cm}, we adopt $u = {\rm log} \, \nu$ or $u = 1+ z$ for the ESMB and 21cm line models, respectively. We work typically with $N_p = 2^9 = 512$. In the case of the redshifted 21cm line, the original grid consists in a number of points slightly smaller than 512, but it can easily be extended considering that the global signal can be set to zero at $z<5$.

Then, we replace $T_{\rm th}^{\rm BB/dist}$ with $T_{\rm th,s}^{\rm BB/dist}$ to work with a smoothed version of the original tabulation. Finally, we proceed to the calculation of the differences $\Delta a_{\ell,0}$ and $\Delta R$ as in Sect. \ref{sec:instability}.

\begin{figure}[h!]
\centering
         \includegraphics[width=8.5cm]{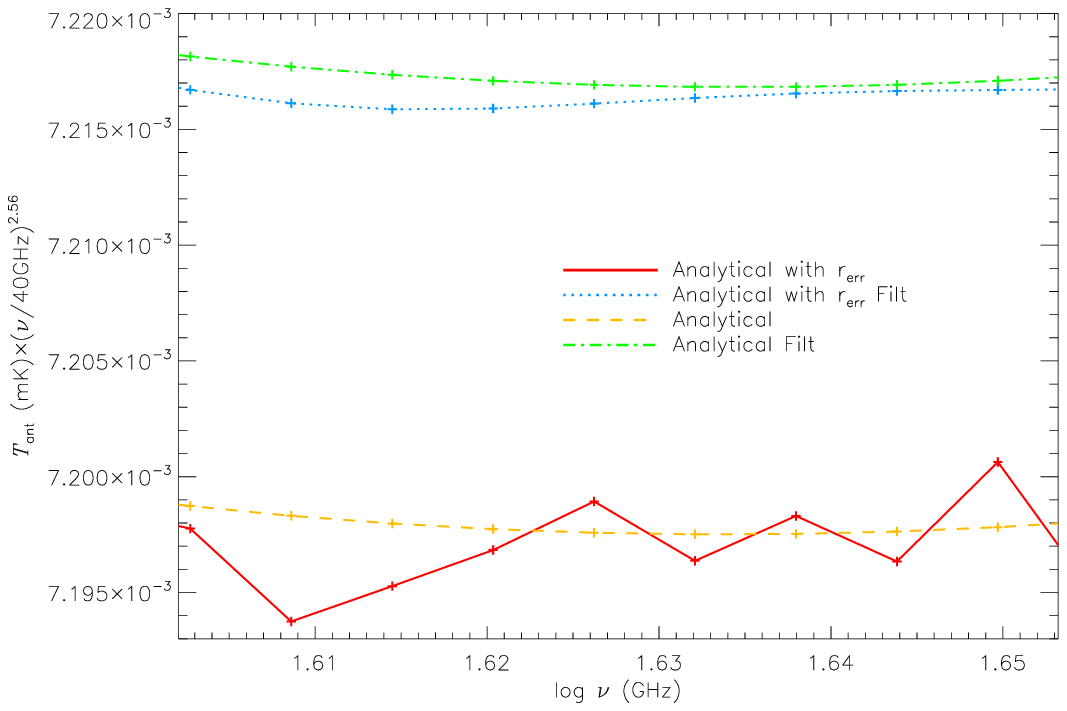}
    \caption{Comparison between the original signals represented by the polynomial interpolations in log-log of the ESMB without any error or adding relative errors and the corresponding filtered ones. We display the signal according to the original tabulation which is equispaced in log$\,\nu$ and multiplying the antenna temperature by a certain power of the frequency to better appreciate small scale effects due to the added noise superimposed to the general trend. We consider a narrow frequency interval corresponding to 9 grid points, as those adopted in the differential scheme to numerically compute subsequent derivatives (Sect. \ref{sec:filt}). See legend and text.}
    \label{fig:PrefSignals}
\end{figure}

\begin{figure}[h!]
\centering
         \includegraphics[width=8.5cm]{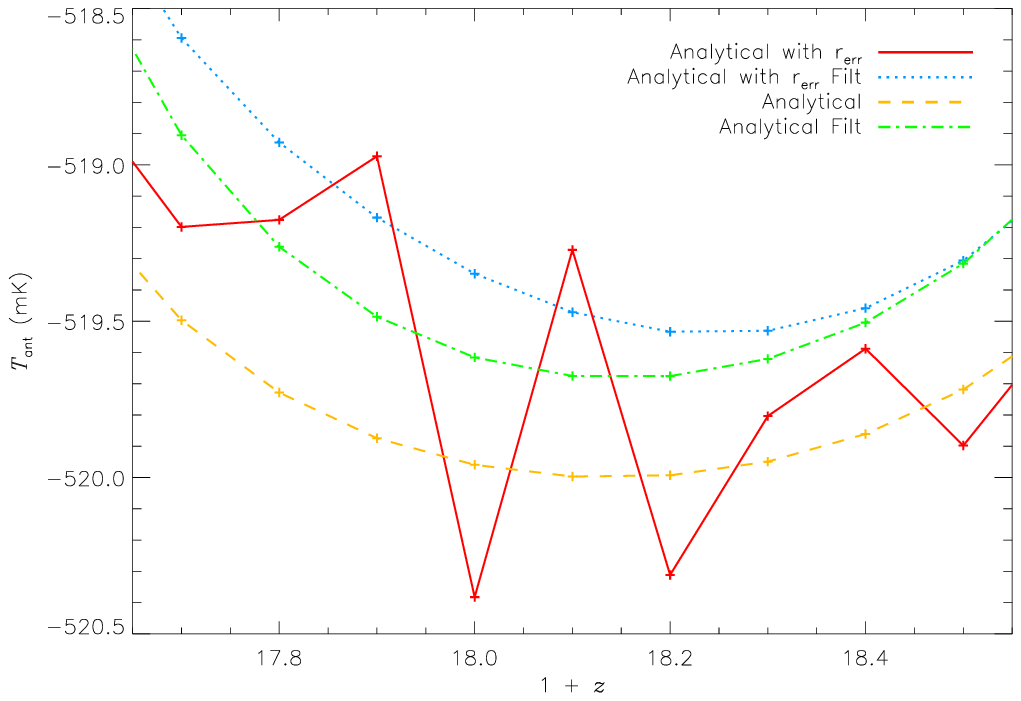}
    \caption{The same as in Fig. \ref{fig:PrefSignals} but for the original signal represented by the EDGES analytical profile. The signal is displayed according to the original tabulation which is equispaced in $1+z$. See legend and text.}
    \label{fig:PrefSignals_21cm}
\end{figure}

Figures \ref{fig:PrefSignals} and \ref{fig:PrefSignals_21cm} show two examples of the effect of the pre-filtering applied with $f=0.1$ to the ESMB and to the redshifted 21cm line in very narrow intervals  of the equispaced variable $u$ to make the effect clearly visible given the large variation of the signal on a wide range of $u$. We compare the analytical reference models without any error or adding errors extracted from a Gaussian random realisation, as discussed in Sect. \ref{sec:instability}, and the corresponding filtered ones. As emerges from the figures, the pre-filtering produces a small change in the original monopole spectrum.
For the ESMB it is less than $\simeq 0.3$\% in the considered frequency range (see Fig. \ref{fig:Back_mm_Int}), while for the 21cm line, as one could expect because of its sharp profile, the relative change keeps always less than $\simeq 5$\%, being of $\simeq 1$\% on average, in the frequency range $\nu \in [66, 90]$ MHz, centred around 78 MHz, where the HI absorption profile is significant.

The most remarkable effect is instead the clear smoothing of the fluctuations introduced by the noise, that will result into a significant mitigation of the instability in the calculation of the differences $\Delta a_{\ell,0}$ and $\Delta R$ (Sect. \ref{sec:results}).

\subsection{Filtering of derivatives}
\label{sec:filt_der}

A second (or subsequent) approach is based on the filtering of the derivatives computed using the centred approximation numerical derivative scheme given in \cite{1988MC}. We considered three different methods.

\subsubsection{Filtering in real space}
\label{sec:filt_real}

For each $u_i$ grid point we initially compute the first derivative at all the $n_{\rm ds}$ contiguous points, $u_j$, as in the \cite{1988MC} scheme. We then apply a low-pass Gaussian filter\footnote{We also tested a moving average low-pass filter but, in this context, it turned out to be a little worse.} in real space to smooth out the first derivative in the $u_i$ point which is estimated as an average over the $n_{\rm ds}$ points $u_j$ with weights

\begin{equation}\label{eq:GRfilt}
G_{ij} \propto {\rm e}^{-\frac{1}{2}\left({\frac{u_j-u_i}{\sigma_G}}\right)^2} \, ,
\end{equation}

\noindent
where $\sigma_G$ determines the level of smoothing, which increases with $\sigma_G$. Possible suitable values of $\sigma_G$, of about $(0.1 - 0.2)\, /\, \sqrt{8\,\rm{ln}\,2}$, are suggested by the considered problems. Indeed, a step of $0.1 - 0.2$ in log$\,\nu$ corresponds to a full width half maximum (FWHM) bandwidth in frequency of few tens of percent, which is comparable with that usually adopted in microwave experiments, while the tabulation used here for the redshifted 21cm line models has a step of 0.1 in $1+z$. 

After that, we compute the second derivative for each point as the first derivative of the filtered first derivative as evaluated above, and again smooth it out. We iterate this scheme up to the sixth order derivative. 
At this point, we can evaluate the filtered derivative of any order at the desired frequencies, i.e. at the set of $\nu'_{\beta,\perp}$ following the derivative scheme (Sects. \ref{sec:al0_deriv}--\ref{sec:scal_deriv}), using the weights in \cite{1988MC} for the zero order derivative, i.e. interpolating over the grid of filtered derivatives. 
In practice, just the weights for the derivative of orders zero and one are used in this method.

Theoretically, it is possible to choose different $\sigma_G$ values at each iteration in order to optimise the filter according to the amplitude of the fluctuations occurring at a given derivation step. In some cases, we tested this option without finding a clear significant benefit, so we decided to keep the $\sigma_G$ value fixed for all the steps.

In principle, this real space filter can also be applied as pre-filtering of the tabulated monopole frequency spectrum, replacing the method described in Sect. \ref{sec:prefilt}, but in this context the results are a little worse.

\subsubsection{Filtering in Fourier space}
\label{sec:filt_FFT}

As an alternative to the filtering in real space, we can filter the derivatives in the Fourier space. For all the points, $u_i$, of the adopted grid we first compute the derivative using the scheme in \cite{1988MC}.
We then smooth it out using the same method described in Sect. \ref{sec:prefilt}, but applied to the derivatives with order larger than zero. 
In principle, the choice of $f$ can also be adjusted according to the order of derivative, but, again, we do not find a clear significant advantage from this.

For this approach, we implemented two different methods.

\begin{itemize}
\item
{\it Filtering sequentially}

\noindent
As in Sect. \ref{sec:filt_real}, we compute the first derivative, then we smooth it out as in Sect. \ref{sec:prefilt}; we then evaluate the second derivative as the first derivative of the first derivative filtered in Fourier space, and again smooth it out; we iterate this scheme up to the sixth derivative. As before, just the weights for the derivative of orders zero and one are used here.

\smallskip

\item
{\it Filtering at once}

\noindent
We compute the derivatives from order zero to order six using the corresponding weights. Then, for derivative orders larger than zero, we independently smooth each derivative out using the filter in Sect. \ref{sec:prefilt}.

\end{itemize}

\noindent
So far, we have computed these filtered derivatives on the adopted grid. Then, we evaluate the filtered derivative at the desired frequencies, i.e. at the set of $\nu'_{\beta,\perp}$ as required by the derivative scheme, using the weights in \cite{1988MC} for the zero order derivative, i.e. for interpolating over the grid of filtered derivatives.

\subsection{Boosting amplification and deamplification}
\label{sec:filt_amp_deamp}

As anticipated in Sect. \ref{sec:intro} and shown in Sect. \ref{sec:instability}, the instabilities in the direct calculation of the $a_{\ell,0} (\nu,\beta)$ coefficients mainly come from the uncertainty in the monopole frequency spectrum at small scales combined to the necessity to perform a fine comparison of the very small differences between signals at very close frequencies, i.e. of the order of the Doppler shift in frequency, $\delta \nu$, which is much smaller than the resolution of the tabulation grid. At the same time, in general and in particular for feature-rich spectral shapes, it is not possible to significantly degrade the adopted grid resolution in order to not spoil the available information about the spectrum shape. Conversely, for a hypothetical observer having a much higher speed, by a significant factor $f_a$, than the real one with respect to the frame at rest with the considered background, the frequency Doppler shift would be correspondingly larger. This implies a comparison between signals at frequencies that, consequently, differ much more than in the case of the observer real speed, potentially decreasing the relative impact of uncertainties. 

The above considerations suggest to investigate the possibility of performing the calculation of the $a_{\ell,0} (\nu,\beta)$ coefficients assuming an amplified speed value, $\beta_a = f_a \beta$ (obviously keeping $\beta_a < 1$), and then to properly rescale them to the real value of $\beta$. To this aim, we can exploit the scaling rules between derivatives given in Sect. \ref{sec:scal_deriv}. In principle, larger values of $\beta_a$ result into larger differences between the relevant frequencies. 
In practice, for each considered observational frequency, $\nu$, we need to work locally according to the available tabulation grid, or, in other words, we need to avoid too much large values of $f_a$ able to move the relevant frequencies outside the range identified by the set of points (typically, $n_{\rm ds} = 9$) used in the differentiation scheme, to retain as much information about the spectrum shape.
From Eq. \eqref{eq:nuboost} we have

\begin{equation}\label{eq:wboost}
w = {\rm cos}\, \theta = \frac{1}{\beta} \left[{ 1 - (1-\beta^2)^{1/2} \frac{\nu'}{\nu}  }\right] \, .
\end{equation}

The condition $|w| \le 1$ implies
\begin{equation}\label{eq:cond}
\frac{1 - \beta}{(1-\beta^2)^{1/2}}  \le \frac{\nu'}{\nu} \le  \frac{1 + \beta}{(1-\beta^2)^{1/2}} \, ;
\end{equation}
\noindent
of course, both $\nu' > \nu$ (blueshift) and $\nu' < \nu$ (redshift) are permitted. 

Instead, we are here interested to find the implications for $\beta$ (in the range $0 < \beta < 1$) coming from the system of the two conditions $w \le 1$ and $w \ge -1$ in order to estimate suitable values of $\beta_a$. After few algebra, they give, respectively,
\begin{equation}\label{eq:soldisw1}
\beta \ge \beta_L = \frac{\nu^2-\nu'^2}{\nu^2+\nu'^2} \,
\end{equation}
\noindent
and
\begin{equation}\label{eq:soldisw-1}
\beta \ge \beta_R = \frac{\nu'^2-\nu^2}{\nu^2+\nu'^2} \, ,
\end{equation}
\noindent
where, clearly, the former is not trivial only for $\nu' < \nu$, the latter only for $\nu' > \nu$.

As discussed above, we need to work locally around $\nu$. Among the $n_{\rm ds}$ frequency points used in the differentiation scheme, let us consider two frequencies, $\nu' = \nu_L < \nu$ equal to the lowest frequency and $\nu' = \nu_R > \nu$ equal to the highest one. We consider the system of the two conditions \eqref{eq:soldisw1} and \eqref{eq:soldisw-1} respectively for $\nu_L$ and $\nu_R$. Its solution, i.e. the strongest of the two conditions, depends on the type of grid adopted. For the tabulations used in this work, having an equispaced step in log\,$\nu$ or in $1+z$, the step in $\nu$ increases with frequency, then $|\nu_R -\nu| > |\nu_L -\nu|$ and the strongest condition corresponds to \eqref{eq:soldisw-1}. We then chose

\begin{equation}\label{eq:solw-1}
\beta_a = f_a \beta = \beta_R(\nu,\nu'=\nu_R) = \frac{\nu_R^2-\nu^2}{\nu^2+\nu_R^2} = \frac{(\nu_R+\nu)  (\nu_R-\nu) }{\nu^2+\nu_R^2} \, .
\end{equation}

\noindent
We note that this does not necessarily imply to work with fixed $f_a$ (or $\beta_a$) for the whole considered frequency range. On the contrary, the value of $f_a$ to be used in this approach is determined by the kind and resolution of the tabulation grid. For example, for an equispaced grid in log\,$\nu$ the value of $f_a$ is constant for all the grid points while for an equispaced grid in $1+z \propto 1/\nu$, $f_a$ increases with $\nu$. 
In general, the value of $f_a$ increases for decreasing resolution (or increasing step).

It is easy to check that setting $\beta = \beta_a$ and $\nu' = \nu_R$ in Eq. \eqref{eq:wboost} we obviously have $w=-1$.
We write $\nu_R = \nu + \Delta$ and $\nu_L = \nu - \Delta + \delta$, with $\Delta>0$ and $\delta>0$, where, given the properties of the considered grids, we locally have $\delta$ less than $\Delta$ by a significant factor and $\Delta \ll \nu$.
Thus, setting $\beta = \beta_a$ and $\nu' = \nu_L$ in Eq. \eqref{eq:wboost} we have $w = 1 + 2 (\Delta^2 -\nu\delta -\Delta\delta) / (\Delta^2 + 2 \nu \Delta) \sim 1 - \delta/\Delta$, i.e. a value slightly less than 1.
This shows that this choice of $\beta_a$ allows to work in a range of Doppler shifted frequencies that essentially span the same range $[\nu_L, \nu_R]$ around $\nu$ adopted in the differentiation scheme, while for the real value of $\beta$ they are much more closer to $\nu$.\footnote{In principle, because of the condition \eqref{eq:soldisw-1}, it is possible to work with $\beta_a$ larger than the one in Eq. \eqref{eq:solw-1} and a larger interval in the differentiation scheme, but relaxing the locality. Furthermore, even for the previous methods, more points can be used in the differentiation scheme if a tabulation with a finer grid is available, but at the cost of a corresponding increasing number of operations and not necessarily with a better precision unless the finer tabulation is effectively more accurate.}

We can compute the $a_{\ell,0} (\nu,\beta)$ coefficients for $\beta_a$ using both the interpolation scheme and the derivative scheme (Sect. \ref{sec:InterpVSderiv}).
Equation \eqref{eq:ratio_der} then allows to rescale the result to the real value of $\beta$. In the interpolation scheme, this rescaling can be globally performed applying it as in the case of the leading 
derivative, of order equal to $\ell$, while, in the derivative scheme, this rescaling can be performed for each derivative order.

This methods, specifically designed for the problem under consideration, acts as a filter that, in the presence of numerical uncertainties, can essentially allow to compare the variations of the monopole frequency spectrum under more numerically stable conditions.

\section{Filtering results}
\label{sec:results}

\subsection{Pre-filtering}
\label{sec:pref}

We study the efficiency of the pre-filtering method described in Sect. \ref{sec:prefilt}. We report the results obtained for a level of smoothing characterised by $f=0.1$.

\subsubsection{$\Delta R$}
\label{sec:deltar}

First, we consider the mitigation effect of the pre-filtering on the estimation of the observed monopole, expressed in terms of the difference $\Delta R$ (see Sect. \ref{sec:instability} for its definition).

\begin{figure}[h!]
\centering
         \includegraphics[width=9cm]{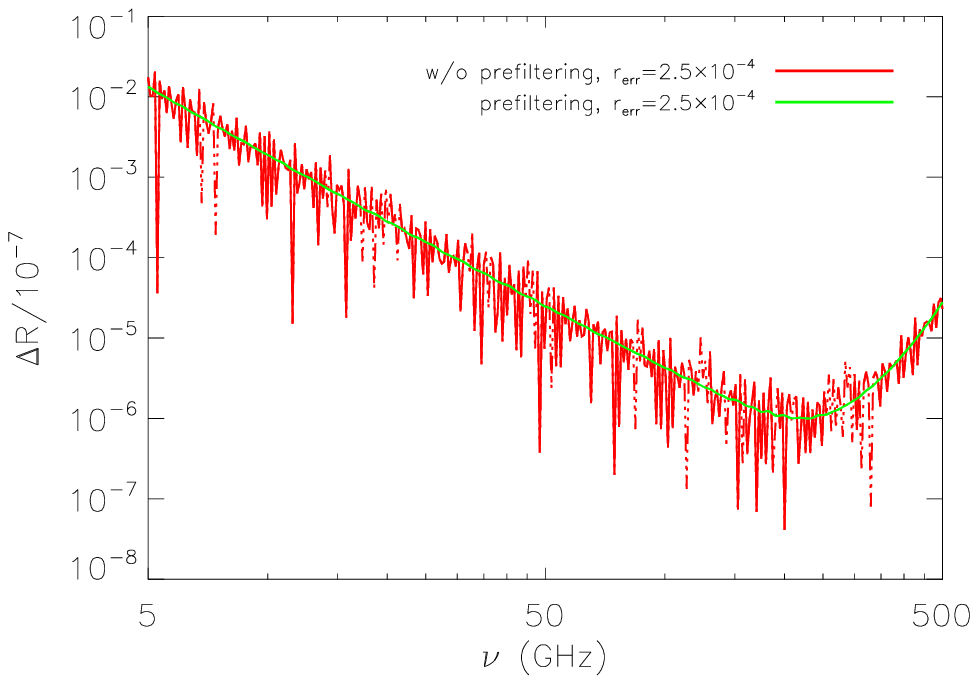}
    \caption{$\Delta R$ evaluated assuming the analytical representation of the ESMB with a detection threshold $S_{max}$ = 0.1 Jy, considering a relative error and including or not the pre-filtering. The curve corresponding to the ideal case is almost superimposed to the green one, thus it is not reported for simplicity. Here (and in the following analogous figures), solid (dots) lines refer to positive (negative) values. See legend.}
    \label{fig:deltaRmm}
\end{figure}

Fig. \ref{fig:deltaRmm} shows this difference assuming the analytical representation of the ESMB signal for the maximum value of the detection threshold and a relative error $r_{\rm err} = 2.5 \times 10^{-4}$, including or not the smoothing. As in principle expected from Eq. \eqref{eq:struct_der_0}, the propagation of the uncertainty in the intrinsic monopole spectrum affects the calculation of the observed one introducing instabilities in the direct calculation that are strongly suppressed by the pre-filtering. 

\begin{figure}[h!]
\centering
         \includegraphics[width=9cm]{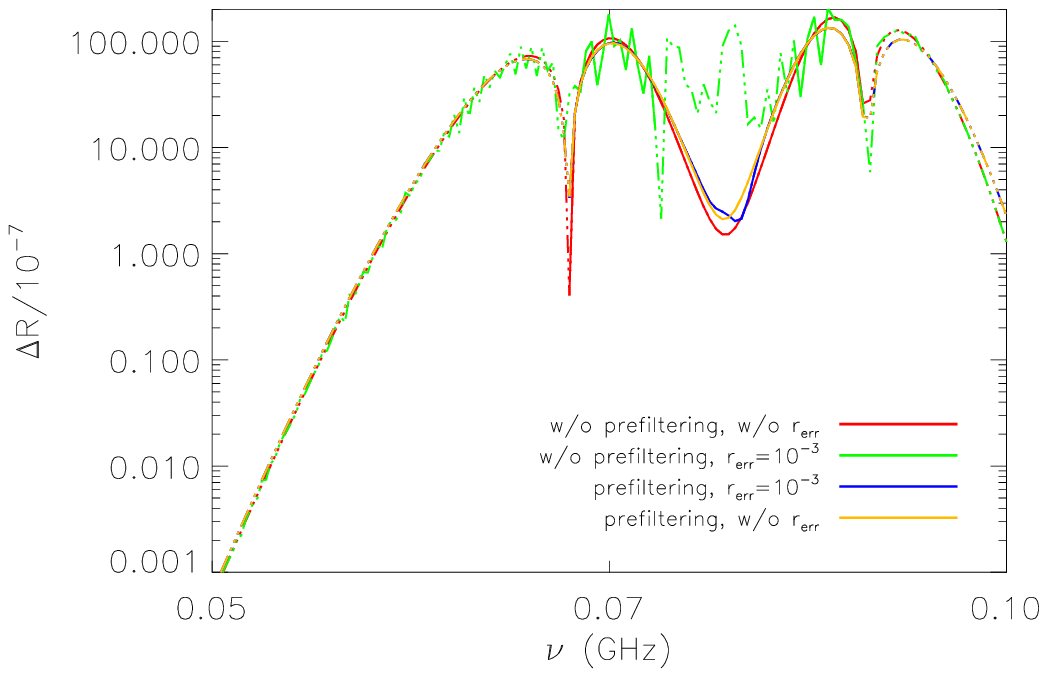}
    \caption{$\Delta R$ evaluated assuming the analytical representation of the EDGES profile including or not a relative error and the pre-filtering. See legend.}
    \label{fig:deltaR21}
\end{figure}

Fig. \ref{fig:deltaR21}, instead, refers to the 21cm line EDGES profile, assuming $r_{\rm err} = 10^{-3}$. Since in this case the pre-filtering effect on the original monopole spectrum is expected to be not so small (Sect. \ref{sec:prefilt}), for comparison we also display the result obtained in the absence of uncertainties. Although we find that the pre-filtering introduces a little smoothing excess at frequencies around the relative minima and maxima of $\Delta R$, it mainly results into a very good mitigation of the artefacts induced by the propagation of numerical uncertainties in the direct calculation.

In the presence of observer motion, this test underlines the relevance of filtering the original tabulated intrinsic monopole spectrum for a stable and accurate estimation of the observed one.

\begin{figure*}[h!]
\hskip -0.6 cm
         \includegraphics[width=10cm]{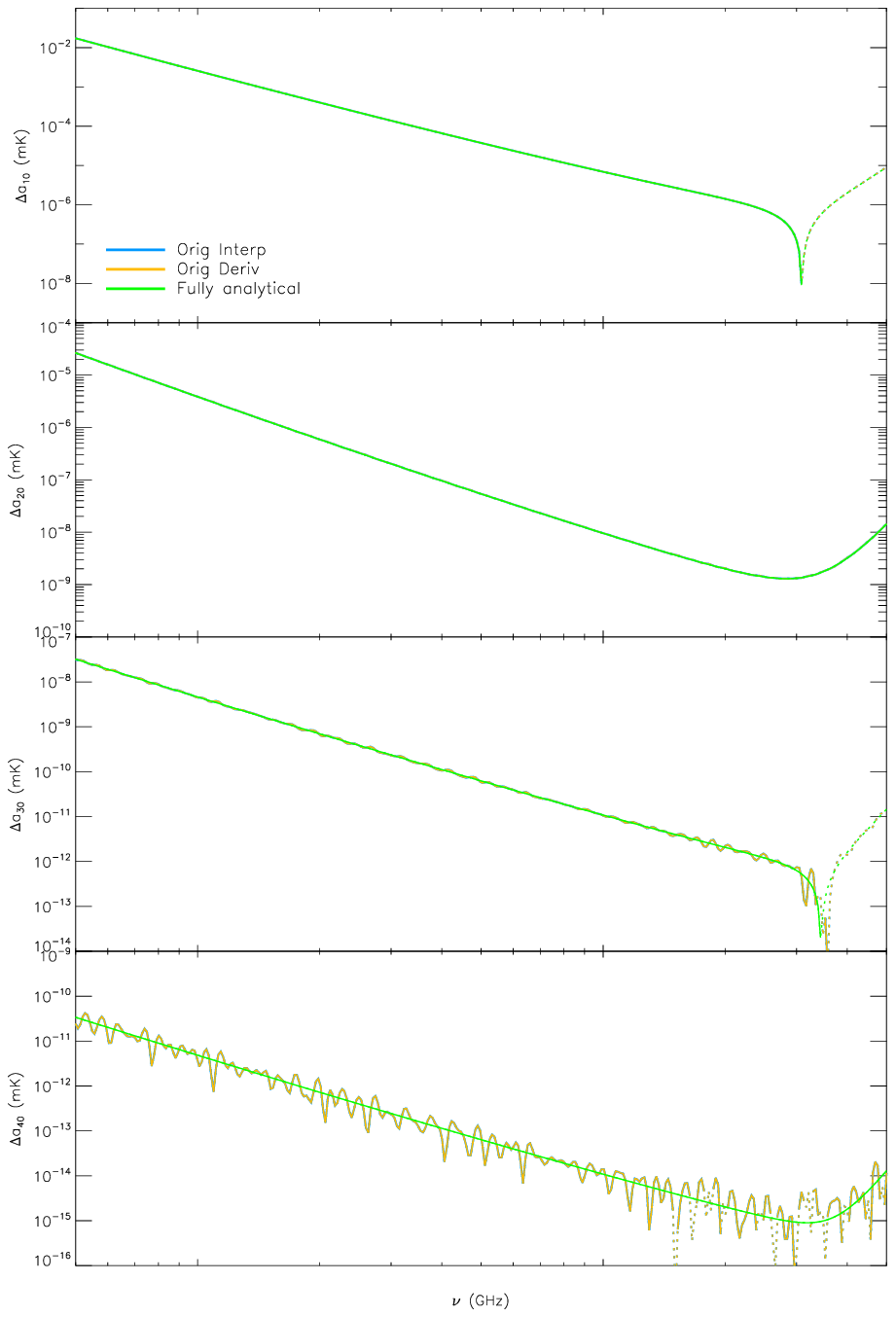}
\hskip -0.7 cm
         \includegraphics[width=10cm]{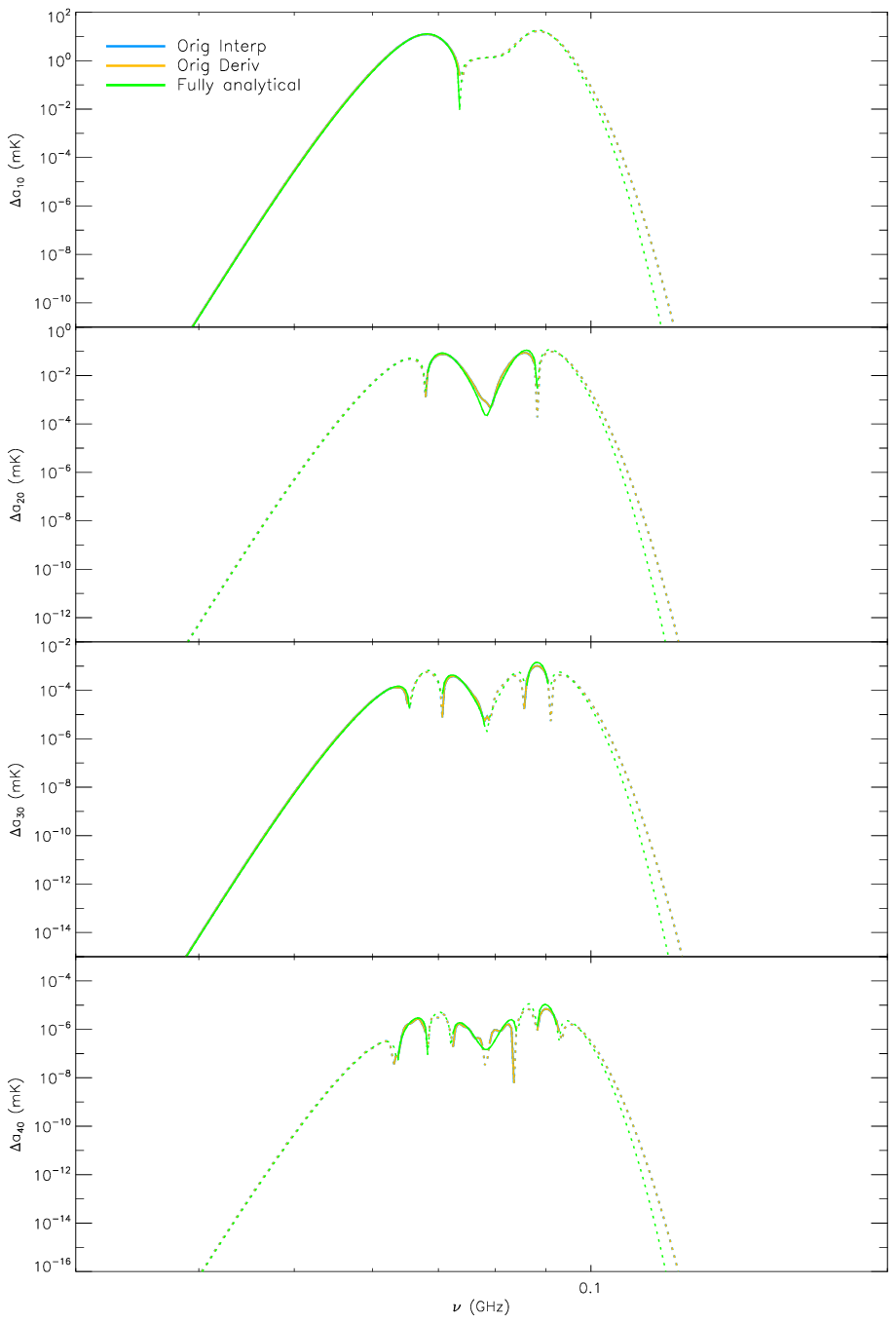}
    \caption{The same as in Fig. \ref{fig:original} but including pre-filtering with $f = 0.1$ and $r_{\rm err}$ as before. See legend.}
    \label{fig:prefiltering}
\end{figure*}

\subsubsection{$\ell \ge 1$}
\label{sec:ellge1}

As emerged in Fig. \ref{fig:original}, the effect of the propagation of numerical uncertainties strongly increases with the multipole, being remarkable even at $\ell = 2$. Here, we focus on the mitigation achieved with the pre-filtering using the interpolation and derivative schemes. The results are shown in Fig. \ref{fig:prefiltering} for the ESMB (left column) and the 21cm EDGES profile (right column), where they are compared with the ideal fully analytical treatment. The method provides a good estimation of the spectrum up to $\ell = 2$, a reasonable mitigation of the instabilities at $\ell = 3$, while it is less effective from $\ell = 4$. The advantage represented by further applying to the signal the filters discussed in Sects. \ref{sec:filt_der} and \ref{sec:filt_amp_deamp} is described in the next sections.

\subsection{Filtering}
\label{sec:filt}

Hereafter, we present the results derived for the different filtering techniques discussed above, separately for the two investigated backgrounds.

\subsubsection{ESMB: analytical model}
\label{sec:app_microw_back}

\begin{figure*}[h!]
\hskip -0.6 cm
         \includegraphics[width=10cm]{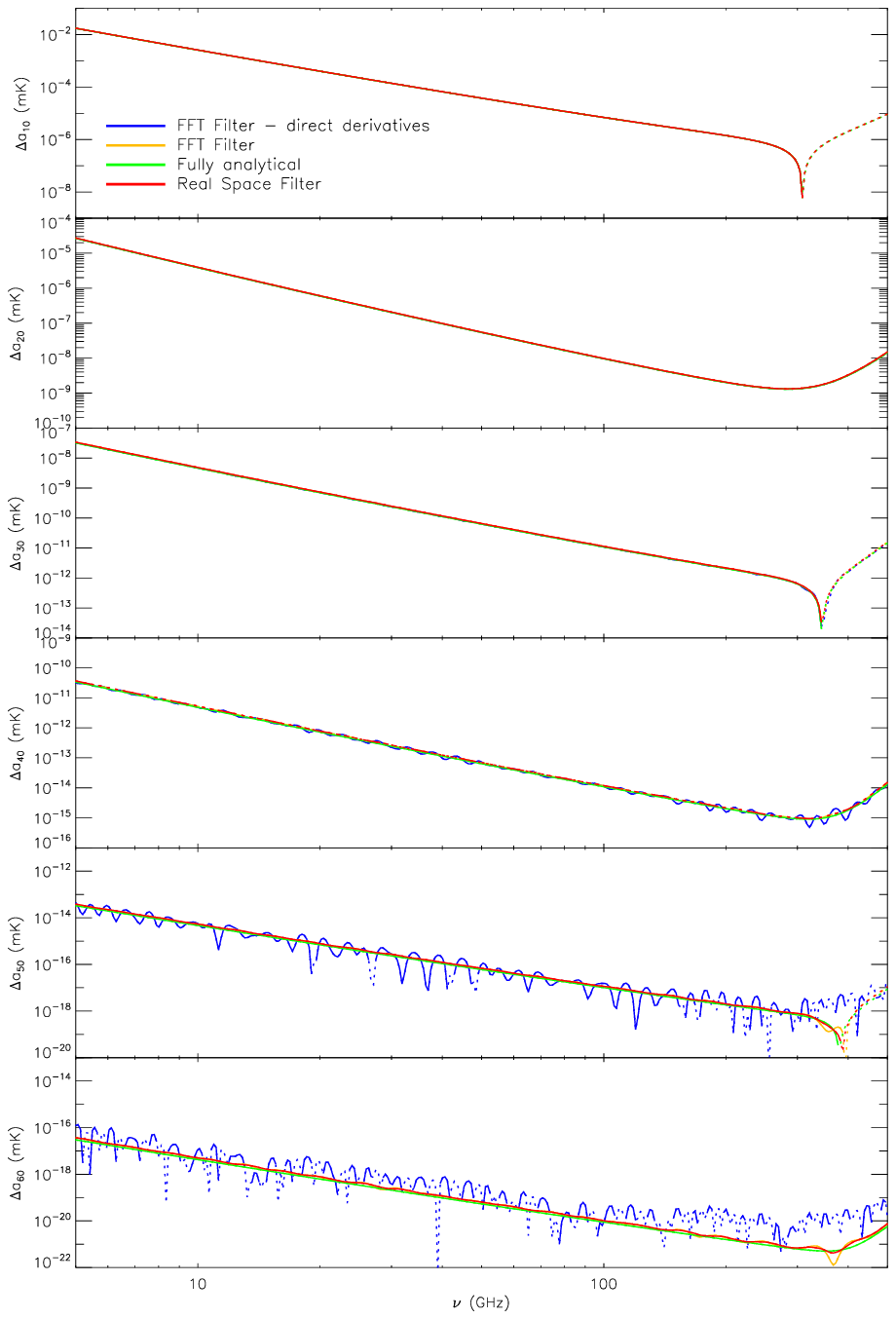}
\hskip -0.7 cm
         \includegraphics[width=10cm]{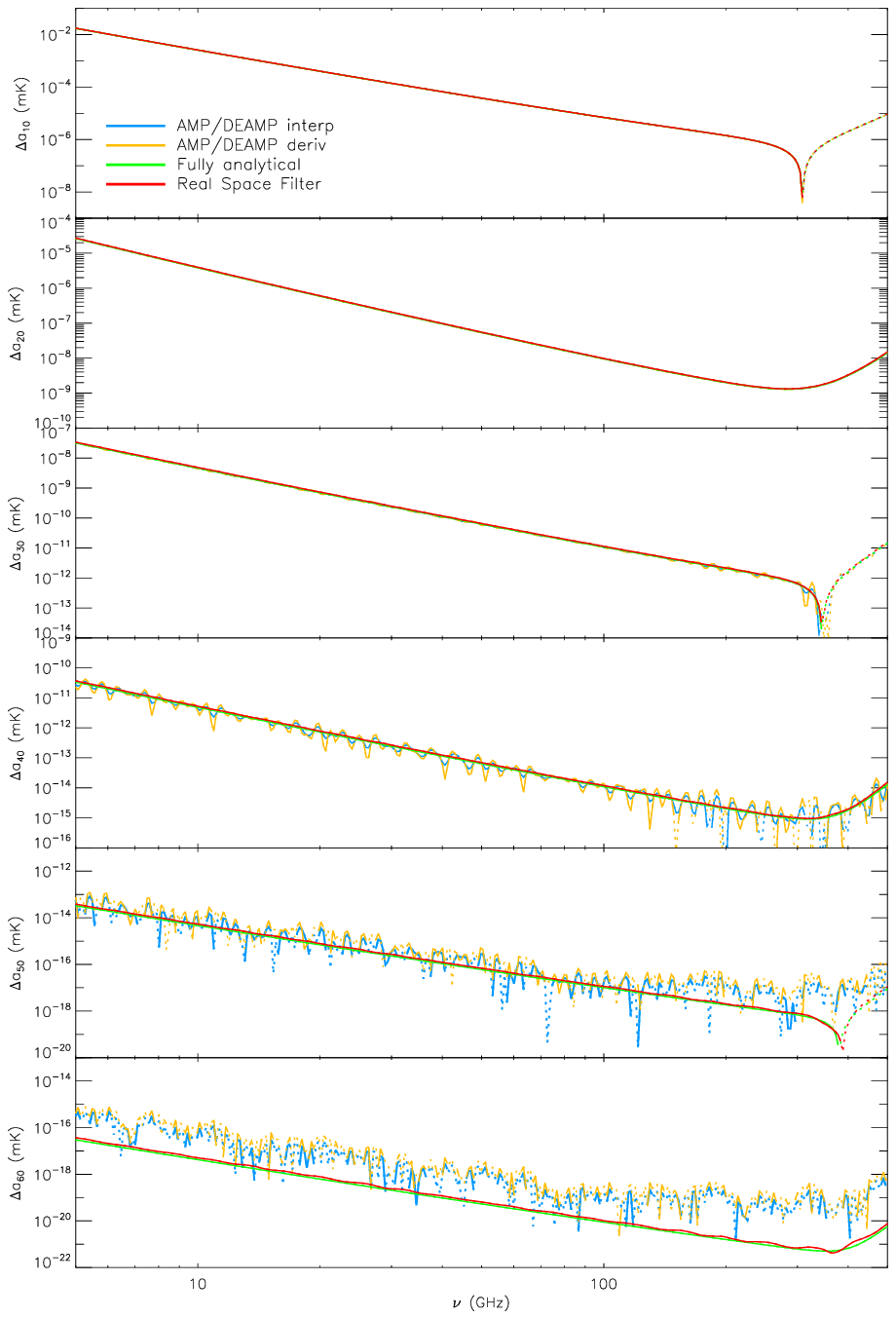}
    \caption{Comparison between the different filtering methods and the ideal fully analytical case, starting from the analytical formulation of the ESMB including the relative error and the pre-filtering as before. See legend and text.}
    \label{fig:filterback}
\end{figure*}

In this section, we compare the spherical harmonic coefficients obtained for the ESMB with the described filters up to $\ell = 6$, starting from the original monopole pre-filtered as described in Sect. \ref{sec:prefilt} with the same parameters as in Sect. \ref{sec:pref}. Specifically, left column of Fig. \ref{fig:filterback} shows the results of the FFT derivative filtering approach applied at once or sequentially, the real space filter and the reference ideal case. These last two curves are also compared with the approach based on boosting amplification and deamplification for both the interpolation and the derivative schemes (right column of Fig. \ref{fig:filterback}). As emerges from the plots and in agreement with the results already obtained applying only the pre-filtering, all the methods well reproduce the spectrum up to $\ell = 2$. Furthermore, although with different efficiency, all of them provide a further mitigation of the instabilities at $\ell = 3$ with respect to the case of pre-filtering alone (Fig. \ref{fig:prefiltering}): this is particularly evident in the case of the filtering of derivatives, in both real and Fourier space, and with boosting amplification and deamplification in the interpolation scheme. The differences between the results obtained with the different filters increase at increasing multipole. At $\ell = 4$, the filtering with boosting amplification and deamplification, particularly in the derivative scheme, and also the FFT derivative filtering applied at once do not produce a significant further mitigation with respect to the case of pre-filtering alone, and their efficiency degrades at increasing $\ell$. On the contrary, filtering in sequence the derivatives allows to significantly mitigate the propagation of the monopole uncertainties up to the maximum multipole investigated, working in both Fourier and real space, the latter resulting more stable than the former even around frequencies where, for odd $\ell$, $\Delta a_{\ell,0}$ changes in sign or, in general, where it assumes low values. 

Appendix \ref{appc} reports few representative cases to illustrate the validity of the method for different uncertainties in the monopole spectrum.

\subsubsection{ESMB: tabulated models}
\label{sec:app_microw_back_tab}

\begin{figure}[h!]
\hskip -0.6cm
         \includegraphics[width=10.cm]{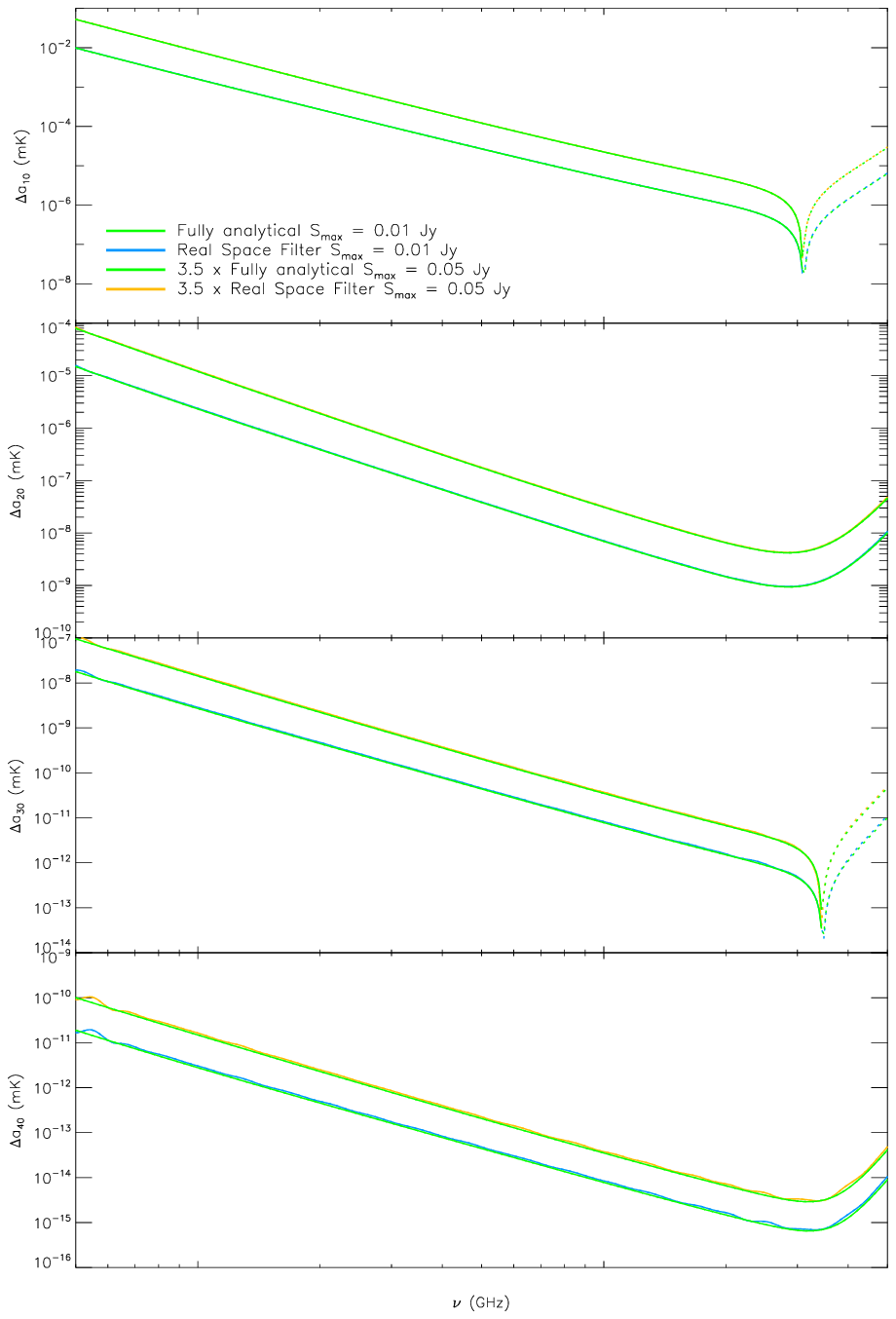}
    \caption{Comparison between the real space filter method applied to the ESMB tabulated intensity calculated adopting two different thresholds and the corresponding ideal fully analytical cases. See legend and text.} 
    \label{fig:totutab2}
\end{figure}

We derived the spherical harmonic coefficients for the ESMB tabular representations under study. As expected, the quality of the predictions does not vary much with the assumed threshold, thus, for simplicity, we report the results only for two different values of $S_{max}$. Fig. \ref{fig:totutab2} shows the spectra up to $\ell = 4$ for $S_{max} = 0.01$ Jy and 0.05 Jy. The latter is multiplied by a factor of 3.5 to better distinguish the curves.
In the figure, pre-filtering is applied. As in the analytical case, the best and more stable filtering approach confirms to be the Gaussian filter in real space, that works very well except at $\ell = 4$ near the low frequency boundary and for very low values of $\Delta a_{4, 0}$ because of the presence of little oscillations.

\subsubsection{Redshifted 21cm line: EDGES}
\label{sec:app_21cmED}

\begin{figure*}[h!]
\hskip -0.6 cm
         \includegraphics[width=10cm]{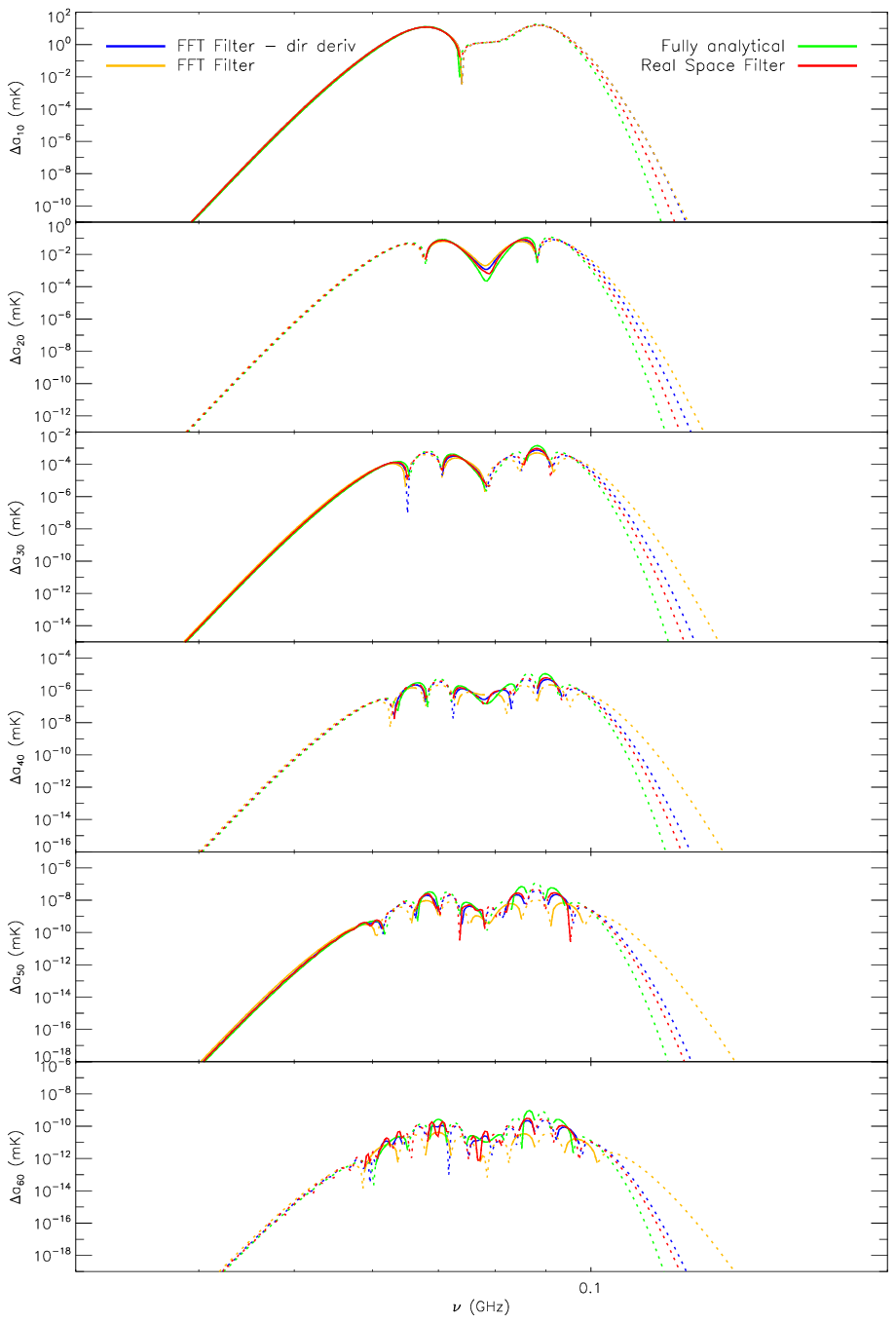}
\hskip -0.6 cm
         \includegraphics[width=10cm]{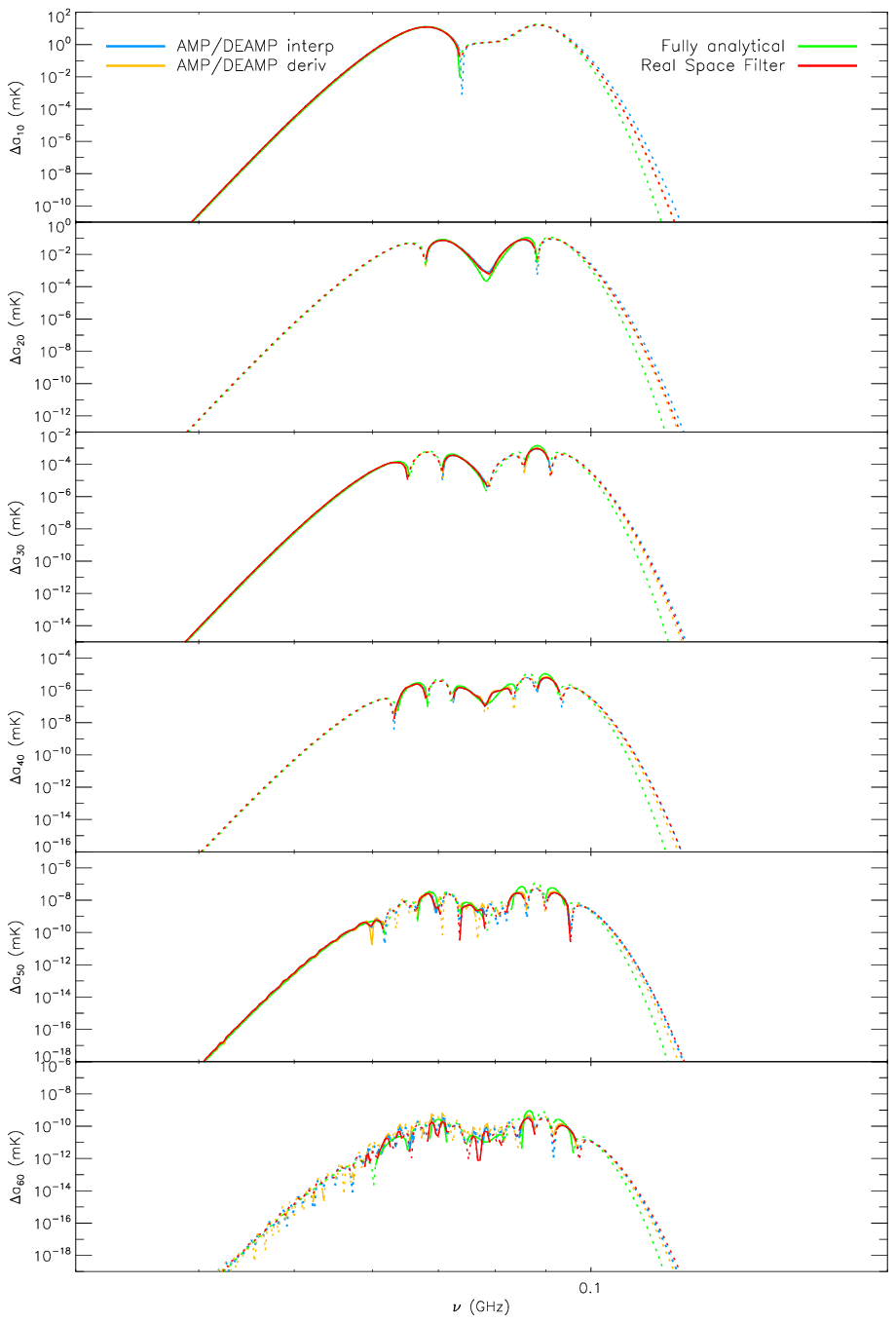}
    \caption{The same as in Fig. \ref{fig:filterback} but for the 21cm EDGES profile. See legend.}
    \label{fig:filter21ed}
\end{figure*}

The improvement in the reconstruction of the spectrum with the above filters, up to $\ell = 6$, is shown in Fig. \ref{fig:filter21ed} for the analytical EDGES profile of the 21cm line. 
In this case, we note at $\ell = 2$ the aforementioned smoothing excess in the spectrum, already found applying only the pre-filtering, that persists around the relative minima and maxima and in the high frequency region. Likely, the latter effect is due to the poorer resolution, $\delta \nu / \nu$, at increasing frequency because of the adopted grid and it is further amplified applying the sequential FFT filter.
In general, the spectrum is well reproduced up to $\ell = 3$, in particular for the real space and the boosting amplification and deamplification filters, and a slight refinement is achieved at $\ell = 4$. At higher multipole these filtering methods allow a reasonable estimation of the spectrum.

In Appendix \ref{appd}, we probe the validity of the method for different uncertainties in the monopole spectrum, focussing, in particular, on the possibility of reducing the smoothing excess induced by the pre-filtering depending on the intrinsic accuracy of the model and the multipole of interest.

\subsubsection{Redshifted 21cm line: tabulated models}
\label{sec:app_21cmtab}

The monopole spectra predicted by the tabulated models described in Sect. \ref{sec:21cm} are very different and cannot be easily characterised by analytical representations; obviously, the same holds at higher multipoles.

\begin{figure*}[h!]
\hskip -0.6 cm
         \includegraphics[width=10cm]{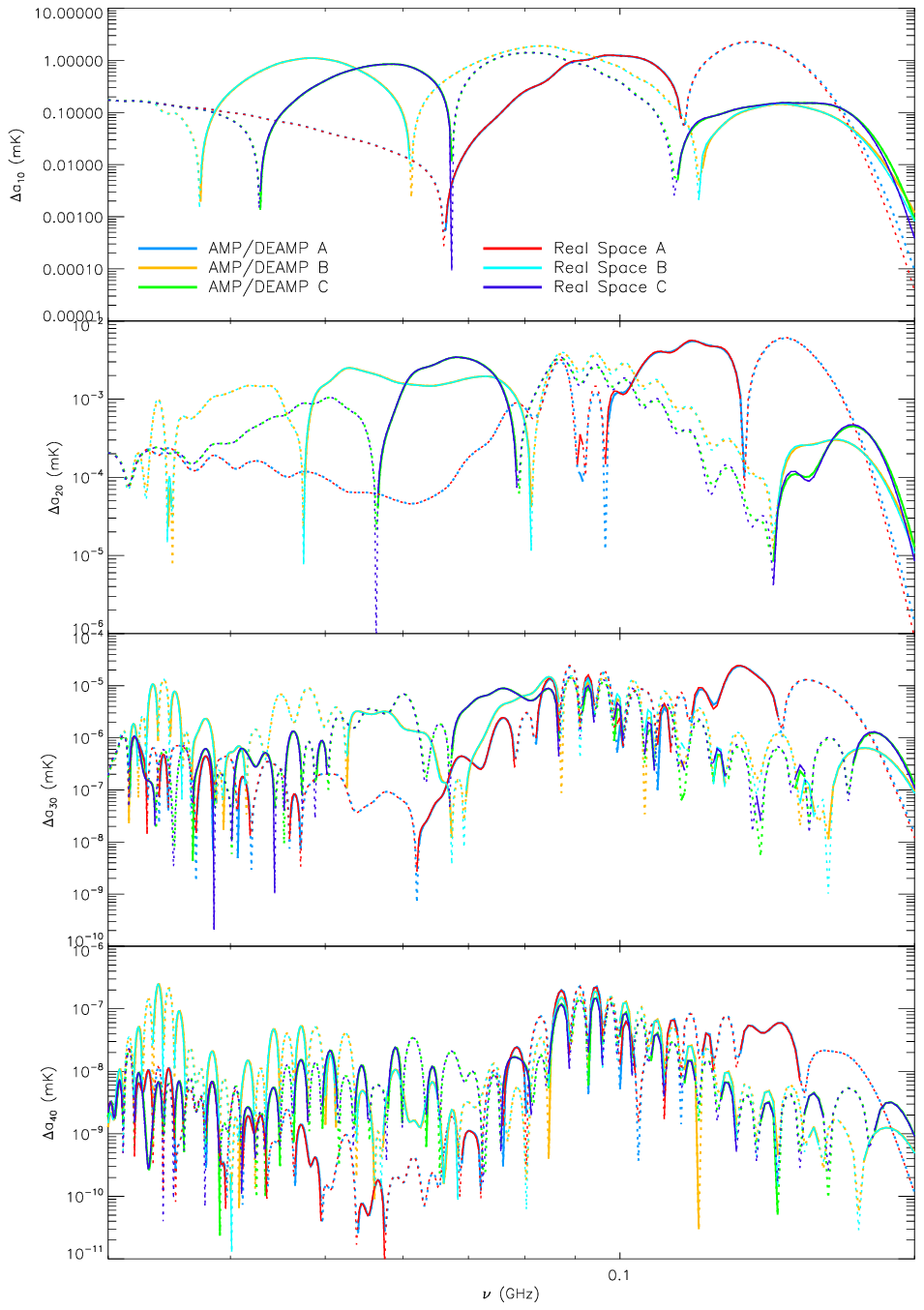}
\hskip -0.7 cm
         \includegraphics[width=10cm]{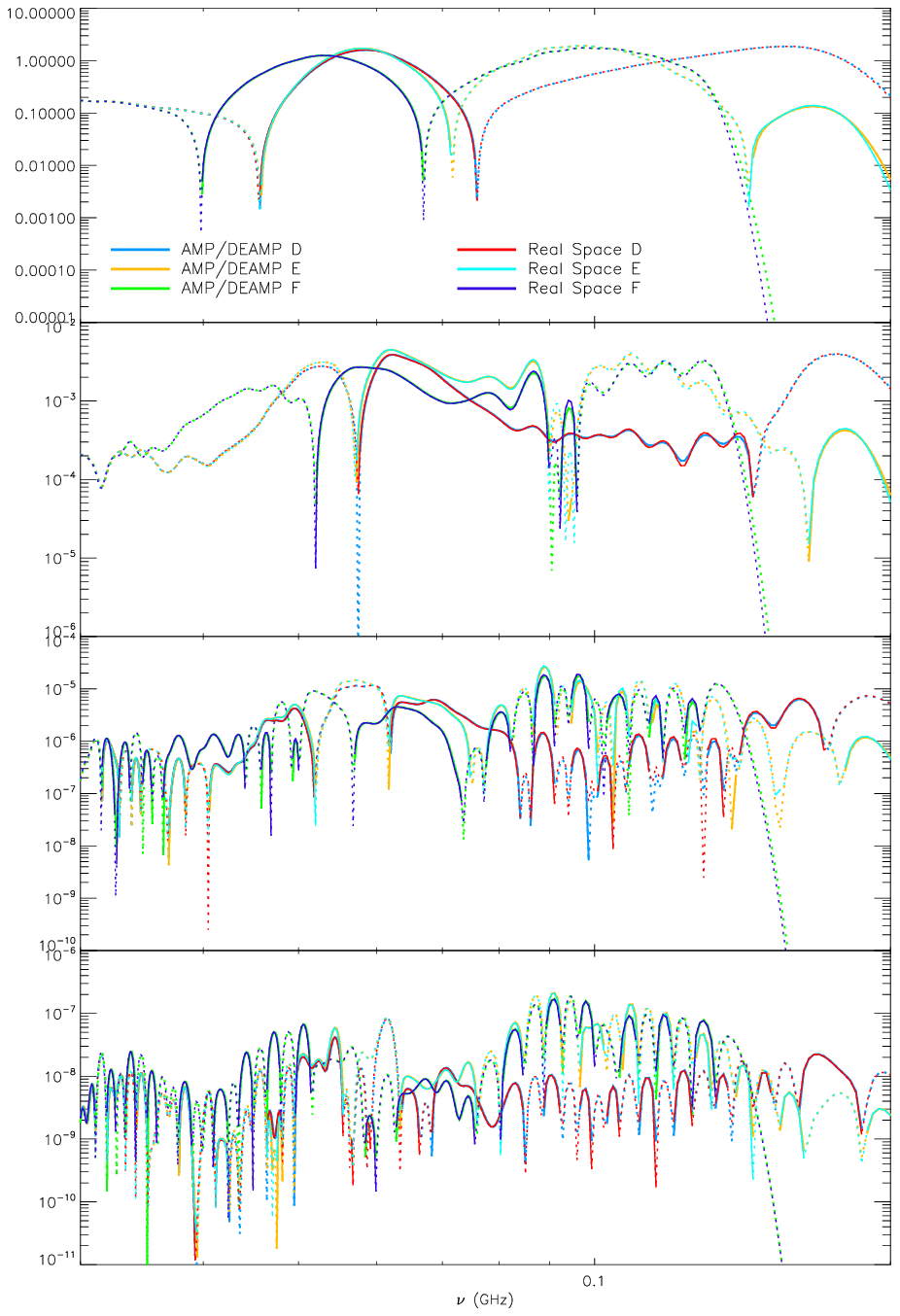}
    \caption{Results for the six considered 21cm line tabulated models applying the pre-filtering and the two best filtering methods, the real space filter and the boosting amplification and deamplification in the interpolation scheme. In order to better distinguish the six models, the $y$-scale differs from that of Fig. \ref{fig:filter21ed}. See legend and text.}
    \label{fig:filter21models}
\end{figure*}

On the other hand, on the basis of the previous results, we are confident that the two best filtering approaches, namely the real space and the amplification and deamplification ones, provide a reasonable prediction of the spectrum up to $\ell = 4$.
As shown in Fig. \ref{fig:filter21models} for the A, B and C models (left column) and for the D, E and F models (right column), the two methods give very similar results.

\section{Discussion and conclusion}
\label{sec:conclu}

In this paper, we studied how to compute the boosting effects produced by the observer peculiar motion, that modifies and transfers to higher multipoles the isotropic monopole frequency spectrum (Sects. \ref{sec:theoframe} and \ref{sec:sol_struct}), beyond the case of cosmic or extragalactic backgrounds well characterised by analytical or semi-analytical functions. Indeed, monopole frequency spectra described by tabular representations are typically affected by uncertainties. If they are negligible, the adopted interpolation or derivative scheme (Sect. \ref{sec:InterpVSderiv}) gives results with accuracy similar to that achieved in the case of analytical or semi-analytical functions (Sects. \ref{sec:probl_microw_back}, \ref{sec:probl_21cm}). Differently, they propagate and relatively increase with the derivative order, possibly preventing an accurate computation of its transfer to higher multipoles and also of the observed monopole (Sects. \ref{sec:probl_microw_back}, \ref{sec:probl_21cm} and \ref{sec:deltar} and Appendices \ref{appa} and \ref{appb}). We developed methods to filter (Sect. \ref{sec:filt}) the original tabulated function or its derivatives and, ultimately, the multipole spectra and identified approaches for deriving reliable predictions for a wide range of background models.

For monopole frequency spectra that can be expanded in Taylor's series, we derived explicit expressions for the harmonic coefficients in terms of monopole spectrum derivatives (Sect. \ref{sec:al0_deriv}), performing our calculation up to $\ell$ = 6. We considered different types of filters: Gaussian pre-filtering in Fourier space of the tabulated function (Sect. \ref{sec:prefilt} and, for applications, Sect. \ref{sec:pref}); Gaussian filtering in real space of the numerical derivatives in sequence (Sect. \ref{sec:filt_real}); Gaussian filtering in Fourier space of the numerical derivatives both in sequence and all at once (Sect. \ref{sec:filt_FFT}); dedicated method of amplification and deamplification of the boosting (Sect. \ref{sec:filt_amp_deamp}) in the interpolation and derivative schemes. We applied them to two very different types of signal usually represented by tabulated functions, namely the (smooth) ESMB (Sect. \ref{sec:microw_back}) and the (feature-rich) redshifted 21cm line (Sect. \ref{sec:21cm}), superimposed to the CMB. We tested these approaches on analytical approximations computed on the adopted tabulation grids and polluted with simulated noise (Sect. \ref{sec:InterpVSderiv}). 

A comparison of the accuracy in calculating the spherical harmonic expansion coefficients using the explicit analytical solutions reported in Sect. \ref{sec:sol_struct} or the standard formal inversion of Eq. \eqref{eq:harm}, that is based on the computation of an integral of the considered monopole background spectrum multiplied by a renormalised associated Legendre polynomial, has already been presented in \cite{2021A&A...646A..75T} (see their equation 5 for the latter method). The formal inversion is performed through a precise quadrature at each $\ell$ that requires the evaluation of the integrand function for a very large number of points (typically from hundreds or thousands or more, depending on $\ell$ and on the spectral shape), while the direct analytic solutions at $0 \le \ell \le 6$ given in Sect. \ref{sec:sol_struct} only require to evaluate the monopole background spectrum in seven points, thus reducing the required computational time by orders of magnitude. Of course, the running time needed for the fully analytical approach depends on the complexity of the functional form, and it is found to be larger for the EDGES profile, $\sim 25$ ms, than for the ESMB, $\sim 3$ ms.\footnote{These times refer to a 2.8 GHz Intel Core i7 (with 16 GB DDR3 RAM); the numerical code has been implemented in Fortran.}

For the various models of the two types of background, the running time is of the order of $\sim 4-5$ ms for pre-filtering, $\sim 200-300$ ms for the boosting amplification and deamplification accounting for both the interpolation and derivative schemes and $\sim 350-450$ ms for the Fourier and real space filters. Without any filtering, applying the direct computation on tabulated models with the interpolation scheme, the running time is $\sim 160-190$\,ms. Since the required execution times are obviously machine dependent, it is more meaningful to provide them in terms of the ratio, $r_t$, between the time required for a computation with a certain type of filtering and that required by the direct explicit solutions in the interpolation scheme. This ratio is of particular interest for backgrounds for which it is difficult to find a suitable analytical representation.
Remarkably, for both the backgrounds, $r_t \sim 1.2-1.3$ for the boosting amplification and deamplification with the two schemes together up to $\sim 2.2-2.3$ for the Fourier or real space filters.

The results found indicate which methods or their combinations are the most appropriate for predicting the spherical harmonic expansion coefficients depending on the background shape, the accuracy of the tabular representation and the multipole of interest.

For many types of smooth background spectra it is feasible to find a set of analytical or semi-analytical functions that, in addition to characterising the monopole spectrum, allow to grasp the main properties of its derivatives with respect to the frequency, or, more in general, of its tiny variations in very narrow frequency ranges. In particular, for the ESMB there is a good agreement between the prediction up to high multipole of a log-log polynomial description and the one derived using a suitable filtering approach. In such conditions, the use of analytical representations is preferable.

On the other hand, feature-rich background monopole spectra and their derivatives are typically difficult to describe with suitable analytical or semi-analytical representations, except for some specific cases. This holds, for instance, for the redshifted 21cm line.

The best approach to manage smooth tabulated background spectra is found to be the combination of pre-filtering in Fourier space of the monopole spectrum with a Gaussian filter of derivatives in real space applied in sequence. This method allowed us to find a robust prediction of the $\Delta a_{\ell, 0}$ differences up to a certain $\ell$, which depends on the intrinsic accuracy of the model. For the accounted reference uncertainty, $r_{\rm err} = 2.5 \times 10^{-4}$, this technique works well up to, at least, $\ell = 4$ (Sects. \ref{sec:app_microw_back} and \ref{sec:app_microw_back_tab}); for uncertainties one order of magnitude smaller (larger) the maximum multipole for a reliable estimate is $\ell = 6$ ($\ell = 3$) (Appendix \ref{appc}). 

When dealing with feature-rich tabulated spectra, such as the 21cm models considered in this paper, we found that pre-filtering in combination with the real space or the boosting amplification and deamplification filtering is the best approach to predict the spherical harmonic coefficients up to $\ell = 4$ (Sects. \ref{sec:app_21cmED} and \ref{sec:app_21cmtab}). For relative uncertainties $r_{\rm err} \lsim 10^{-4}$ we argued that these methods provide good estimates up to $\ell =$ 6 (Appendix \ref{appd}). For such small intrinsic uncertainty of the model and depending on the multipole of interest, the pre-filtering can be avoided in order to reduce the little smoothing excess it might introduce, resulting into an improvement for the dipole and the quadrupole (Appendix \ref{appd}).

The developed methods can significantly extend the range of manageable models, beyond the case of backgrounds characterised a priori by analytical or semi-analytical functions, requiring only an affordable increase in computational time compared to direct calculation performed simply by interpolating the available tabular representations.
 
\appendix

\section{Instability for ESMB tabulated model}
\label{appa}

\begin{figure}[h]
\hskip -0.6cm
         \includegraphics[width=10.cm]{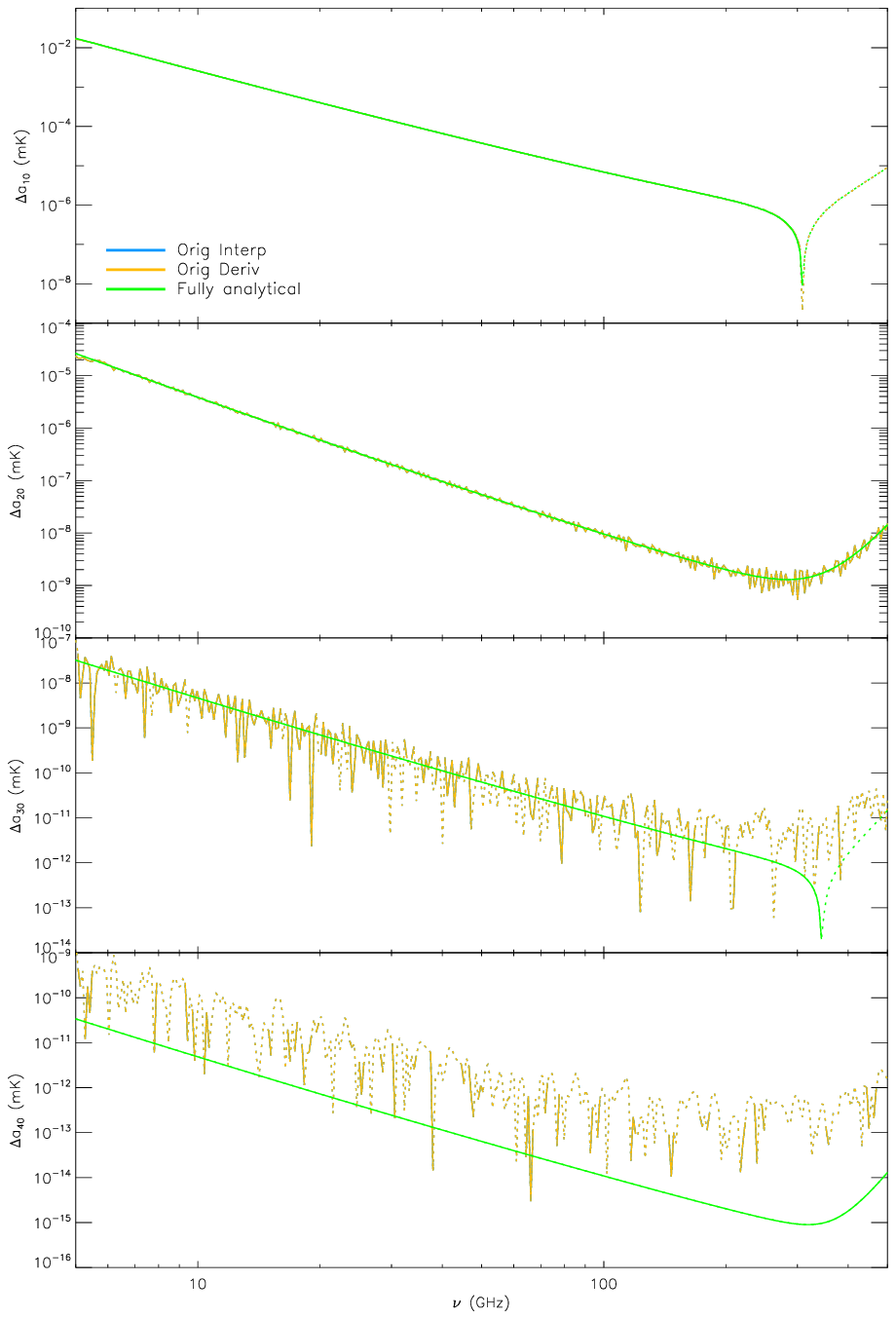}
    \caption{The same as in the left column of Fig. \ref{fig:original}, but for the ESMB described by the original tabulated function. See legend.} 
    \label{fig:totutab}
\end{figure}

Assuming an ESMB described in terms of tabulated function for the largest threshold $S_{max} =0.1$\,Jy (Sect. \ref{sec:21cm}) and using the interpolation and derivative schemes described in Sect. \ref{sec:InterpVSderiv}, we report in Fig. \ref{fig:totutab} the results found applying directly the analytical solutions given in Sects. \ref{sec:sol_struct}--\ref{sec:scal_deriv} and compare them with the ideal fully analytical calculation and the results discussed in Sect. \ref{sec:probl_microw_back} (see Fig. \ref{fig:original}).
As evident from the figure, the intrinsic uncertainties in the original tabulation of ESMB monopole spectrum propagate to higher $\ell$, already at $\ell = 2$, and their effect dramatically increases with $\ell$, making the $\Delta a_{\ell,0}$ computation highly unstable, similarly to the case of the corresponding polynomial representation polluted with numerical uncertainties.

\section{Absolute error instability: EDGES}
\label{appb}

\begin{figure}[h!]
\hskip -0.6cm
         \includegraphics[width=10.cm]{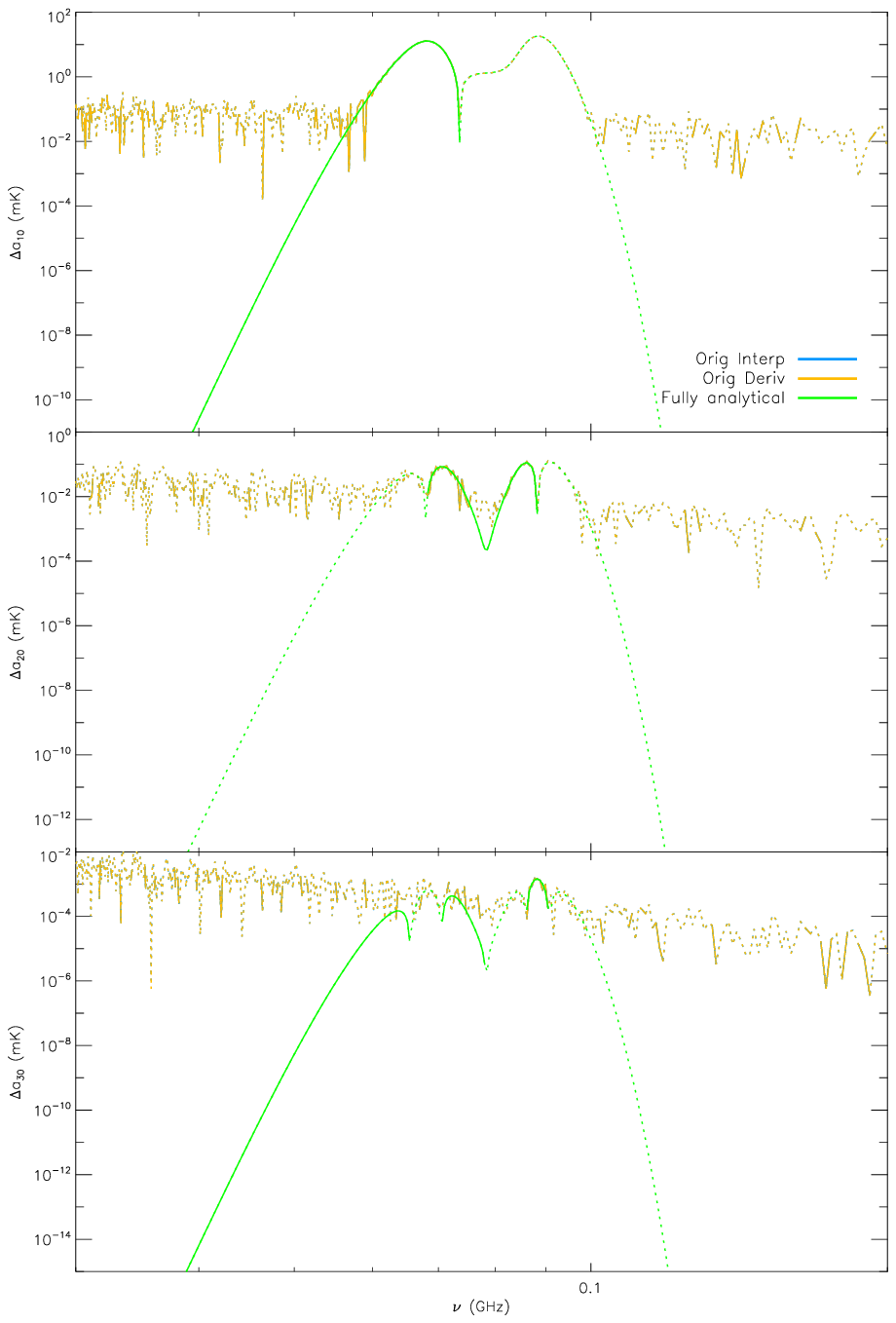}
    \caption{The same as in the right column of Fig. \ref{fig:original}, but for $r_{\rm err} = 0$, $a_{\rm err} = 0.1$\,mK and for $\ell \le 3$. See legend.} 
    \label{fig:21abserr}
\end{figure}

We apply directly the solutions (Sects. \ref{sec:sol_struct}--\ref{sec:scal_deriv}) to the analytical representation of the EDGES absorption profile (Sect. \ref{sec:21cm}), but polluted with inaccuracies independent of the signal and parametrized by a constant absolute error $a_{\rm err}$ (see Eq. \eqref{eq:T&N}).
The results are summarised in Fig. \ref{fig:21abserr} for the interpolation and derivative schemes described in Sect. \ref{sec:InterpVSderiv}. We assume $a_{\rm err} = 0.1$\,mK, a value of about an order of magnitude less than the absolute uncertainty currently quoted, analogously to that of $r_{\rm err}$ adopted in Sect. \ref{sec:probl_21cm}, for a comparison with the results discussed in that section (see Fig. \ref{fig:original}) and with the ideal fully analytical calculation. 

As emerges from the figure, the effect of numerical uncertainties is relatively smaller (larger) at larger (smaller) absolute values of $\Delta a_{\ell,0}$ with respect to the case of inaccuracies proportional to the signal: remarkably, the two positive peaks at $\ell = 2$ and the positive peak between 80\,MHz and 90\,MHz at $\ell = 3$ are now less affected than in Fig. \ref{fig:original}, while numerical uncertainties spoil the quality of signal prediction at small absolute values of $\Delta a_{\ell,0}$.

\section{Uncertainty dependence: ESMB}
\label{appc}

\begin{figure}[h!]
\hskip -0.6 cm
         \includegraphics[width=10cm]{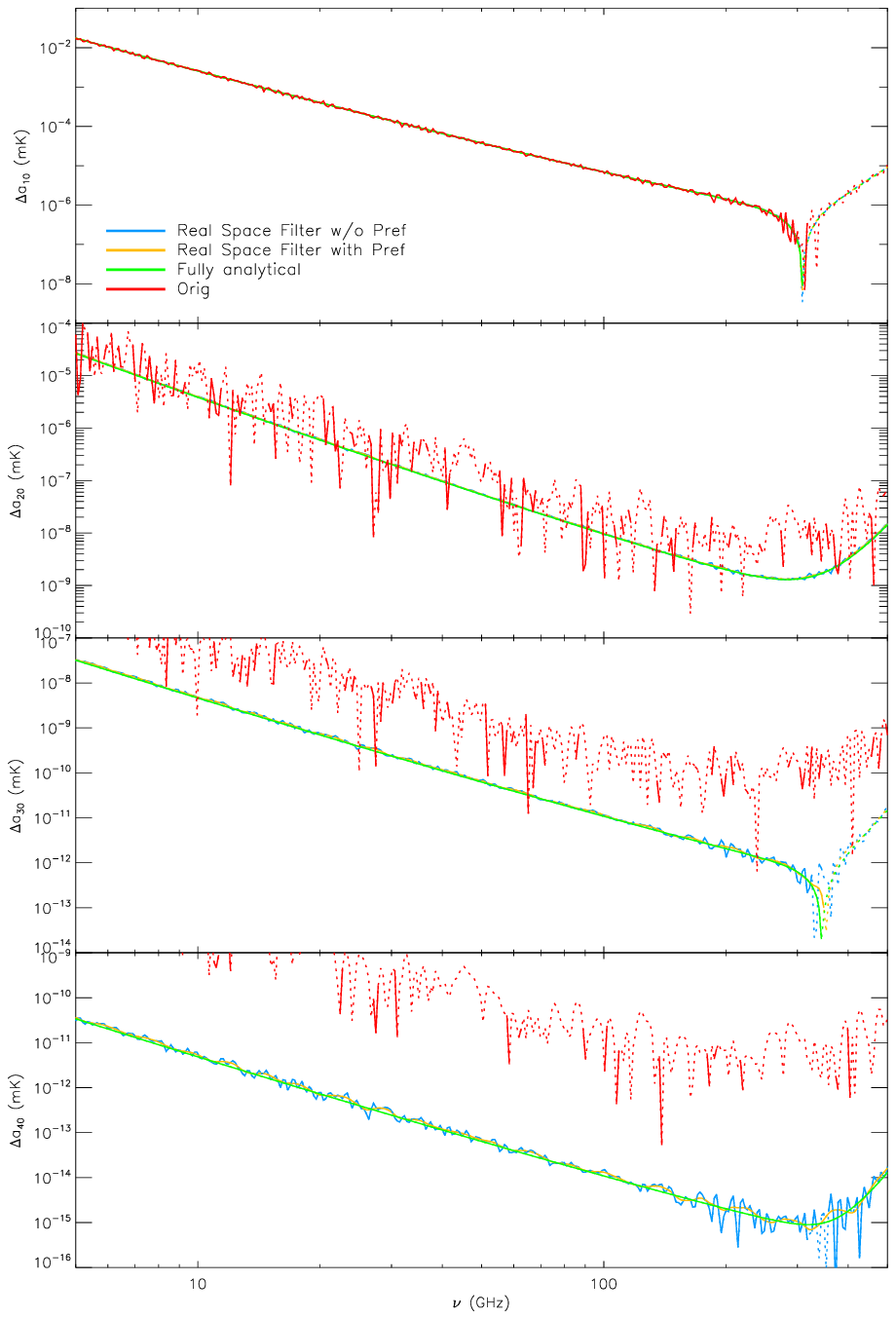}
    \caption{Comparison between the real space filter without and with pre-filtering, the original and the ideal fully analytical cases, starting from the analytical formulation of the ESMB for $r_{\rm err} = 2.5 \times 10^{-3}$ and $S_{max} = 0.1$ Jy. See legend and text.} 
    \label{fig:embbigerrr}
\end{figure}

We investigated how different values of the relative error impact the spectrum prediction. For comparison with previous cases, we considered the ESMB analytical model with $S_{max} = 0.1$ Jy. Fig. \ref{fig:embbigerrr} shows the spherical harmonic expansion coefficients up to $\ell = 4$ for $r_{\rm err} = 2.5 \times 10^{-3}$, one order of magnitude greater than the reference error adopted in this work, applying or not the pre-filtering and compare them with the fully analytical approach and the original treatment. As emerges from the figure, without applying any filter, the coefficients present non negligible oscillations already at $\ell = 1$, that increase at higher multipoles. They are significantly mitigated up to $\ell = 2$ just with the real space filter only, while the pre-filtering plays an important role in damping the oscillations at $\ell \ge 3$. 

\begin{figure}[h!]
\hskip -0.2 cm
         \includegraphics[width=7cm, angle=90]{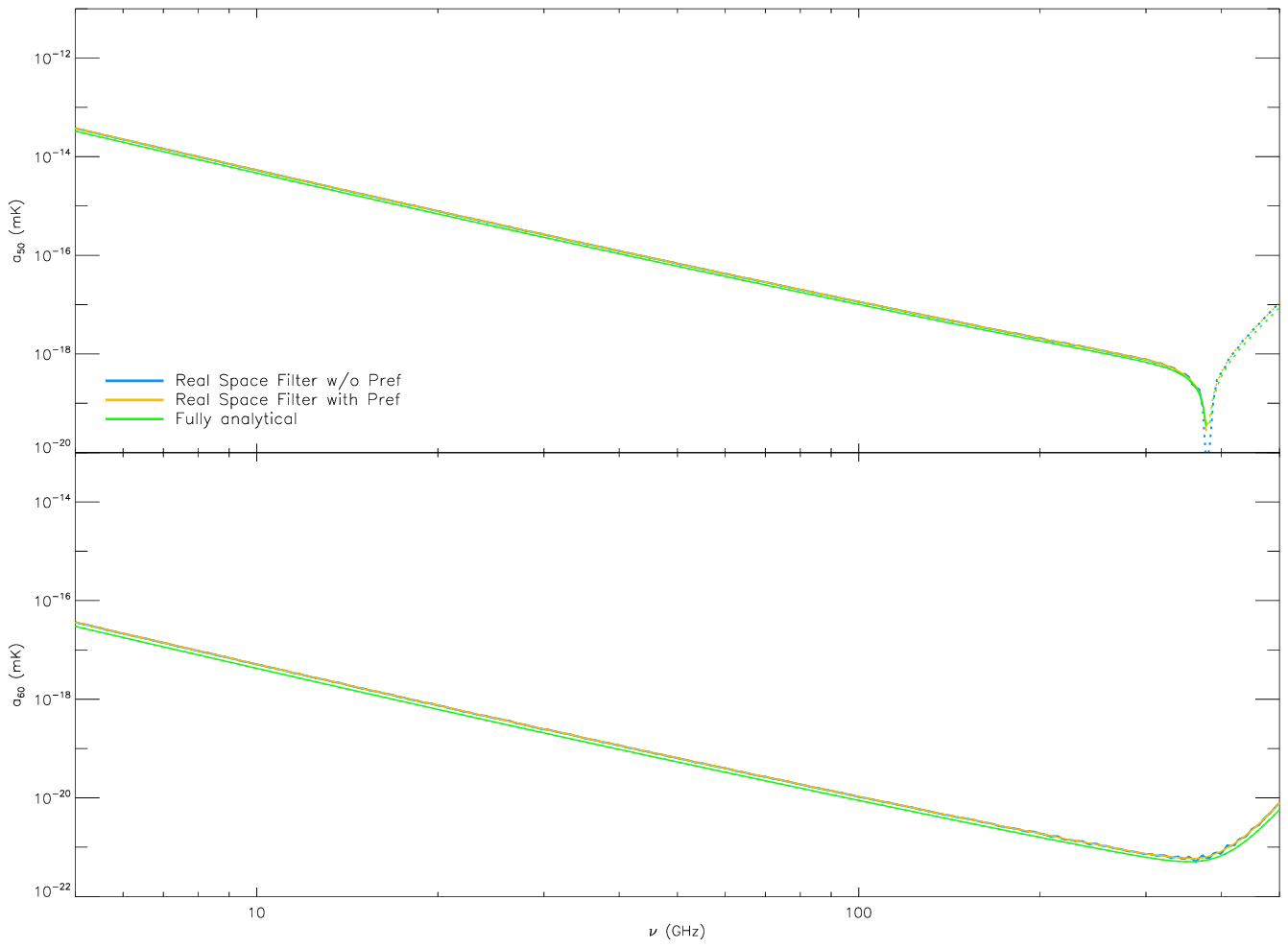}
    \caption{The same as Fig. \ref{fig:embbigerrr} but without the original case, for $r_{\rm err} = 2.5 \times 10^{-5}$ and for $\ell = 5$ and 6. See legend.} 
    \label{fig:embsmallerrr}
\end{figure}

As expected, by decreasing the value of the error to $r_{\rm err} = 2.5 \times 10^{-5}$, the resulting coefficients are much more stable. For this reason, only the last two highest multipoles are shown in Fig. \ref{fig:embsmallerrr}. Here, the spectrum is very well reproduced also without the pre-filtering step, even though with increasing oscillations at low signal values. Compared with the reference case, we note a little increase of the predicted $\Delta a_{\ell, 0}$ values, especially at $\ell = 6$. 

\section{Uncertainty dependence: EDGES}
\label{appd}

\begin{figure}[h!]
\hskip -0.6cm
         \includegraphics[width=10.cm]{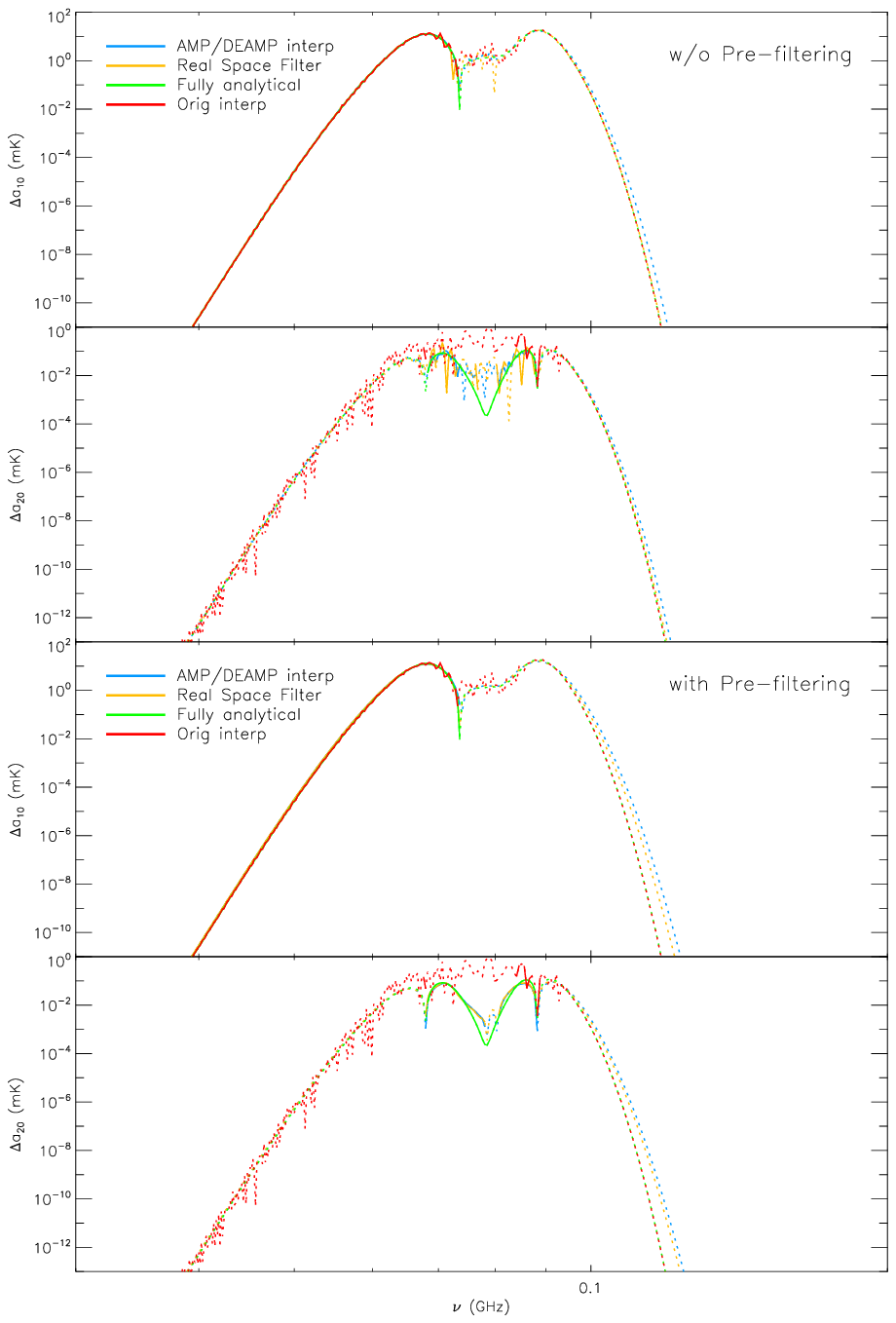}
    \caption{Comparison between the two favourite filters including (two bottom panels) or not (two top panels) the pre-filtering, the original and the ideal fully analytical cases for the EDGES profile for $r_{\rm err} = 10^{-2}$. See legend.} 
    \label{fig:21cmbigerr}
\end{figure}

Analogously to Appendix \ref{appc}, we study the impact of different relative uncertainties on the prediction of the spectra for the analytical EDGES profile. As before, we first consider $r_{\rm err} = 10^{-2}$, an order of magnitude greater than the reference value previously assumed. In Fig. \ref{fig:21cmbigerr} we show the results, up to $\ell = 2$, with and without the pre-filtering for the two best filters and compare them with the ideal and the original cases. As already evident from Sec. \ref{sec:prefilt}, the pre-filtering attenuates the fluctuations, particularly for the real space filter for the dipole ($\ell = 1$), and it becomes more relevant in the reconstruction of the expected spectral shape for the quadrupole ($\ell = 2$) for both filters.

\begin{figure*}[h!]
\hskip -0.6 cm
         \includegraphics[width=10cm]{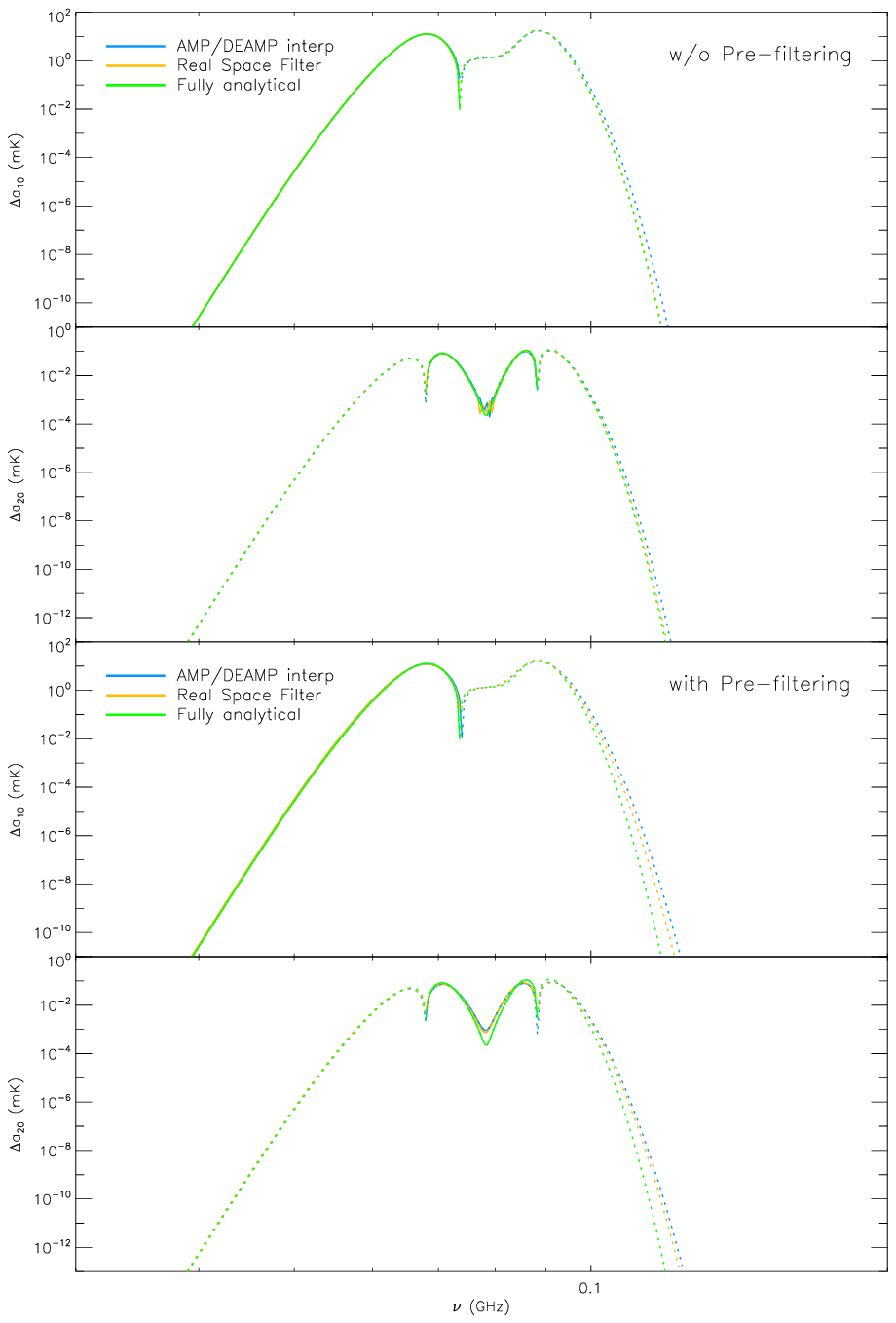}
\hskip -0.7 cm
         \includegraphics[width=10cm]{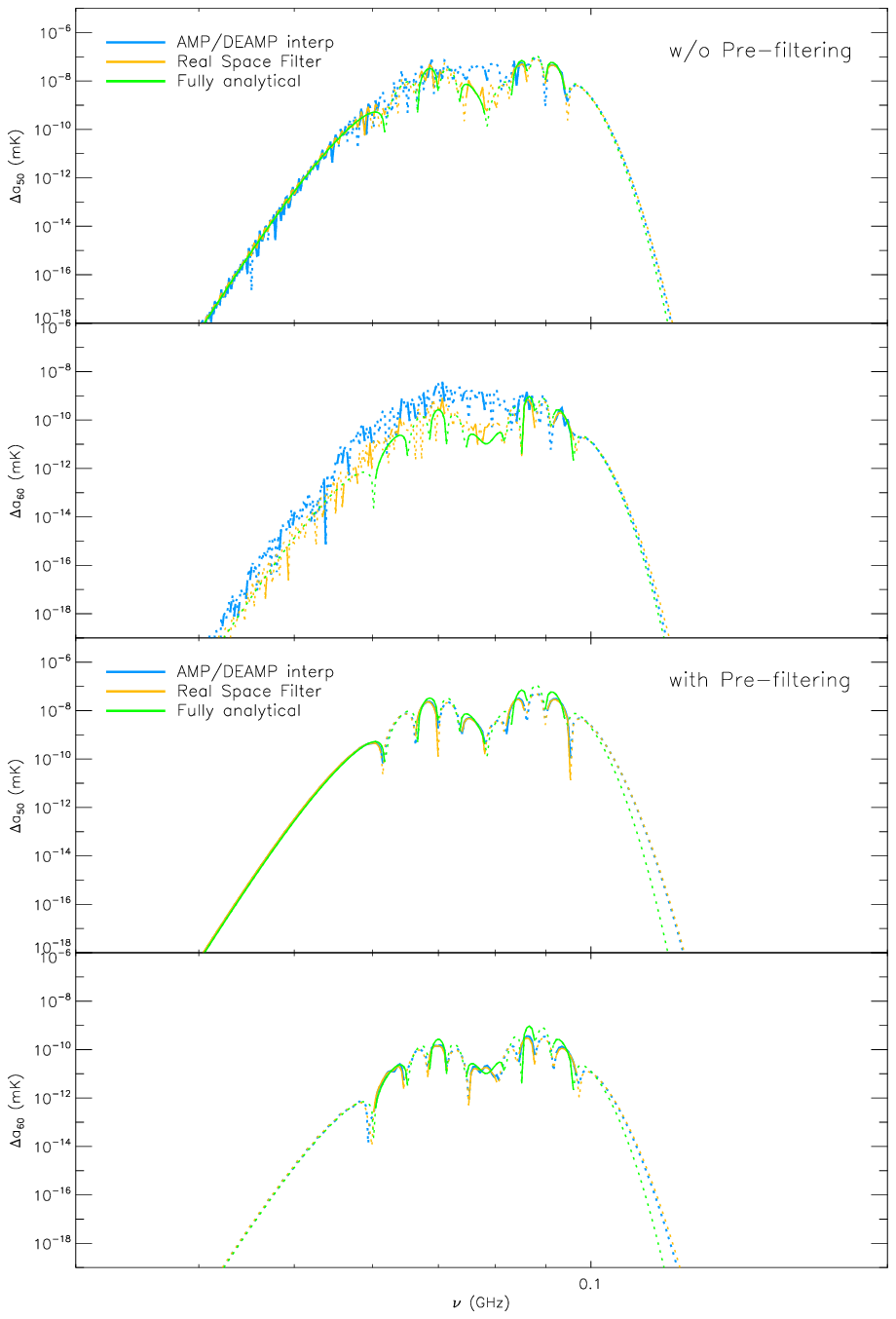}
    \caption{The same as Fig. \ref{fig:21cmbigerr} but without the original case, for $r_{\rm err} = 10^{-4}$ at $\ell = 1$ and 2 (left column) and at $\ell = 5$ and 6 (right column). See legend.}
    \label{fig:21cmsmallerr}
\end{figure*}

On the contrary, for a significantly smaller uncertainty, the dipole spectrum is very well reproduced without the pre-filtering but applying the real space filter or the boosting amplification and deamplification, as evident from the comparison between the first and the third panel in the left column of Fig. \ref{fig:21cmsmallerr}, where $r_{\rm err} = 10^{-4}$. The same conclusion holds for the quadrupole (compare second and fourth panels in the left column), where the two positive maxima better agree with the ideal case, even though the relative minimum around 80 MHz exhibits some little oscillations. Instead, for the spectrum prediction at the highest multipoles, see the right column of the same figure, the mere use of one of the two above best filters does not substantially attenuate the instabilities, while this is possible when these filters are preceded by pre-filtering, which therefore represents a fundamental step.

\section{Contribution from higher order derivatives}
\label{appe}

\begin{figure}[h!]
\centering
         \includegraphics[width=8.5cm]{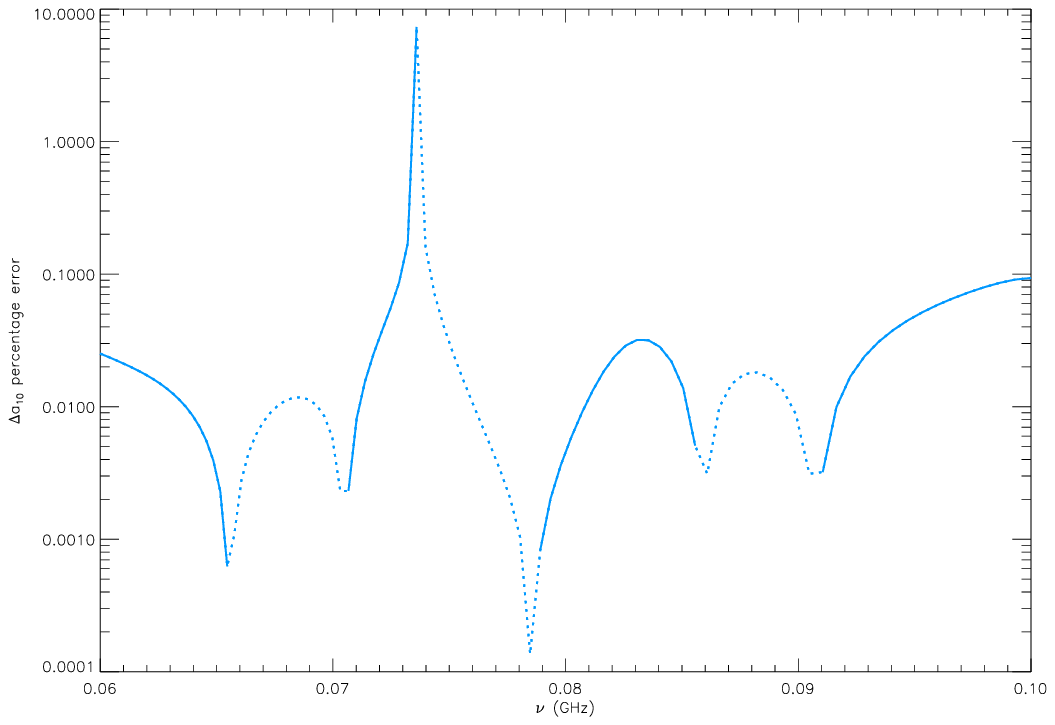}
    \caption{Percentage difference between $\Delta a_{1,0}$ computed for the analytical description of the EDGES absorption profile using Eq. \eqref{eq:struct_der_1} with all the terms and taking only the contribution from the first order derivative, in the ideal case without errors. We note that the expected divergence at a frequency around $\simeq 73.5$ MHz disappears only due to frequency sampling. Solid (dots) lines refer to positive (negative) values. See text.}
    \label{fig:1orderr}
\end{figure}

Let us return to the concept, anticipated at the end of Sect. \ref{sec:al0_deriv}, focussing on the dipole and on the basis of the found results. Considering Eq. \eqref{eq:struct_der_3} at the leading order, we have $T_{\rm th}^\dthree \simeq 15 a_{3,0} / \sqrt{4\pi/7}$. Inserted in Eq. \eqref{eq:struct_der_1} taken up to the third order derivative, this implies $a_{1,0} \simeq {\tilde a}_{1,0} + 2.3 a_{3,0}$, where ${\tilde a}_{1,0}$ includes only the first order derivative contribution to $a_{1,0}$.
The differences $\Delta a_{\ell,0}$ have different frequency dependences at different $\ell$, and, in particular, they exhibit sign changes at frequencies that are different at each $\ell$ \citep{2021A&A...646A..75T}. Thus, in the frequency ranges around these sign inversions, neglecting the terms from higher order derivatives clearly implies significant relative errors in the $\Delta a_{\ell,0}$ calculation even at $\ell = 1$. For example, in the case of the 21cm line, the comparison between the values of $\Delta a_{1,0}$ and $\Delta a_{3,0}$ in the right panel of Fig. \ref{fig:original} shows that, at frequencies around $\simeq 73.5$ MHz, neglecting the contribution from $T_{\rm th}^\dthree$ implies relative errors on $\Delta a_{1,0}$ above the per cent level, i.e. well above the level quoted on simple scaling rules. Fig. \ref{fig:1orderr} shows the relative contribution to $\Delta a_{1,0}$ from derivatives of order greater than one. Similar considerations hold for other types of background and, obviously, this issue is exacerbated when considering combinations of signals with different amplitudes and spectral behaviours for which it is a priori unknown the frequency range where this effect occurs. On the contrary, including higher order terms in the calculation makes this issue negligible.

\begin{acknowledgements}
LT acknowledges the Spanish Ministerio de Ciencia e Innovaci\'on for partial financial support under the projects PID2022-140670NA-I00 and PID2021-125630NB-I00. It is a pleasure to thank the anonymous referee for comments that helped improve the paper.

\end{acknowledgements}

\end{document}